\documentclass{aa}

\pdfoutput=1
\usepackage{graphicx}
\usepackage{txfonts}
\usepackage{hyperref}
\usepackage{natbib}
\usepackage{nicefrac}
\usepackage{multirow}

\begin{document} 
 
\title{The dynamically young outflow of the Class 0 protostar Cha-MMS\thanks{The reduced datacubes are available at the CDS via anonymous ftp to \href{http://cdsarc.u-strasbg.fr/}{cdsarc.u-strasbg.fr(130.79.128.5)} or via \href{http://cdsarc.u-strasbg.fr/viz-bin/qcat?J/A+A/}{http://cdsarc.u-strasbg.fr/viz-bin/qcat?J/A+A/}
}}


   \author{L.~A.~Busch\inst{1}$^,$\inst{2},
          A.~Belloche \inst{1}, 
          S.~Cabrit\inst{3}$^,$\inst{4},
          P.~Hennebelle\inst{5},
          B.~Commer\c{c}on\inst{6}        
          }

   \institute{Max-Planck-Institut f\"ur Radioastronomie, Auf dem H\"ugel 69, 53121 Bonn, Germany\\
              \email{labusch@mpifr-bonn.mpg.de}
         \and
             Argelander-Institut für Astronomie, Universit\"at Bonn, Auf dem H\"ugel 71, 53121 Bonn, Germany
        \and
             LERMA, Observatoire de Paris, PSL Research University, CNRS, Sorbonne Universit\'{e}, UPMC Univ. Paris 06,
             75014 Paris, France  
         \and
             Universit\'{e} Grenoble Alpes, CNRS, IPAG, 38000 Grenoble, France
         \and
         	AIM, CEA, CNRS, Universit\'{e} Paris-Saclay, Universit\'{e} Paris Diderot, Sorbonne Paris Cit\'{e}, 91191 Gif-sur-Yvette, France
         \and
         	Centre de Recherche Astrophysique de Lyon UMR5574, ENS de Lyon, Univ. Lyon1, CNRS, Universit\'{e} de Lyon, 69007 Lyon, France
         }

   \date{Received ; accepted }

  \abstract
 {On the basis of its low luminosity, its chemical composition, and the absence of a large-scale outflow, the dense core Cha-MMS1 located in the Chamaeleon I molecular cloud was proposed as a first hydrostatic core (FHSC) candidate a decade ago.}
   {Our goal is to test this hypothesis by searching for a slow, compact outflow driven by Cha-MMS1 that would match the predictions of MHD simulations for this short phase of star formation.}
   {We use the Atacama Large Millimetre/submillimetre Array (ALMA) to map Cha-MMS1 at high angular resolution in CO 3--2 and $^{13}$CO 3--2 as well as in continuum emission.}
   {We report the detection of a bipolar outflow emanating from the central core, along a (projected) direction roughly parallel to the filament in which Cha-MMS1 is embedded and perpendicular to the large-scale magnetic field. The morphology of the outflow indicates that its axis lies close to the plane of the sky. We measure velocities corrected for inclination of more than 90~km~s$^{-1}$ which is clearly incompatible with the expected properties of a FHSC outflow. Several properties of the outflow are determined and compared to previous studies of Class~0 and Class~I protostars. The outflow of Cha-MMS1 has a much smaller momentum force than the outflows of other Class 0 protostars. In addition, we find a dynamical age of 200--3000 yr indicating that Cha-MMS1 might be one of the youngest ever observed Class 0 protostars. While the existence of the outflow suggests the presence of a disk, no disk is detected in continuum emission and we derive an upper limit of 55 au to its radius.}
   {We conclude that Cha-MMS1 has already gone through the FHSC phase and is a young Class 0 protostar, but it has not brought its outflow to full power yet.}

   \keywords{stars: formation -- stars: protostars -- ISM: individual objects: Cha-MMS1 -- ISM: jet and outflows        }

    \authorrunning{L.A.~Busch et al.}
   \maketitle
%
\section{Introduction}\label{sect:intro}

Star formation undergoes several stages. In the early phase, an initial pre-stellar core collapses
gravitationally and when thermal pressure is high enough, forms a hydrostatic core in the centre which
is called a Class 0 protostar. However, according to the former theoretical work of 
\citet{Larson69}, there is an additional intermediate state whose central core is larger and
less dense than the subsequent Class 0 protostar. This intermediate phase is referred to as the first
hydrostatic core (FHSC) phase. This first core forms when thermal pressure balances the gravitational
force and for a moment, ceases the collapse with the gas being still molecular. As accretion continues the
central temperature keeps increasing. At a temperature of 2000~K, H$_2$ dissociates which consumes energy
and eventually leads to a second, fast collapse until pressure again balances the gravitational force and
the second core or Class 0 protostar forms, subsequently. Since the FHSC stage is expected to be 
short-lived \citep[$\sim$1000~yr for a strongly magnetized core,][]{Commercon12}, its detection
is rather challenging and until now, only one of the candidates, Barnard~1b-N \citep{Gerin15},
seems to fulfil the requirements predicted by magneto-hydrodynamic (MHD) simulations. The results of such
simulations depend on the level of rotation and the mass-to-magnetic-flux ratio but they generally show
the FHSC to be the driving source of a poorly collimated, low-velocity outflow ($\sim5$~km/s)
\citep[e.g.,][]{Hennebelle08, Machida08, Ciardi10, Commercon10, Tomida10, Machida14}, unlike Class 0 protostars which after formation
launch a highly collimated jet with velocities of $>$30~km/s \citep[][]{Banerjee06, Machida08, Machida14}. Given its low
velocity, the outflow remains very compact at the end of the FHSC stage ($\sim$1000~au for a speed of
5~km/s and a lifetime of 1000~yr). Simulations also predict internal luminosities of 0.1~$L_\odot$
over most of the FHSC's lifetime \citep{Commercon12}. \citet{Gerin15} detected a slow, compact outflow
associated with Barnard~1b-N as the driving source. This detection, along with a dynamical age of $\sim$1000~yr and a
low internal luminosity strongly suggest that Barnard~1b-N is a FHSC. 

\begin{table*}
\caption{Setup of the ALMA observations.}
\label{t:setup}
\centering
\begin{tabular}{cccccccccccccc}
\hline
\hline
\noalign{\smallskip}
Conf. & Date of & $t_{\rm start}$\tablefootmark{a} & $N_{\rm a}$\tablefootmark{b} & Baseline  & $t_{\rm int}$\tablefootmark{d} & pwv\tablefootmark{e} & \multicolumn{3}{c}{Calibrators\tablefootmark{f}} & & \multicolumn{3}{c}{Beam\tablefootmark{g}}\\
\cline{8-10} \cline{12-14}
\noalign{\smallskip}
      & observation & (UTC) &  & range\tablefootmark{c} & & & B & A & P & & maj. & min. & P.A. \\
      & (yyyy-mm-dd) & (hh:mm) & & (m) & (min) & (mm) & & & & & ($\arcsec$) & ($\arcsec$) & ($^{\circ}$) \\
\hline
\noalign{\smallskip}
1  & 2015-12-29 & 08:42 & 40 & 15--310 & 34 & 0.4 & 1 & 2 & 3 & & 1.0 & 0.80 & 9  \\
   & 2015-12-30 & 07:56 & 39 & 15--310 & 34 & 0.6 & 4 & 2 & 3 & & 1.1 & 0.79 & $-$4  \\
2  & 2016-06-16 & 22:37 & 38 & 17--704 & 46 & 0.9 & 1 & 4 & 3 & & 0.51 & 0.30 & 30 \\
   & 2016-06-17 & 23:31 & 36 & 17--650 & 46 & 0.6 & 1 & 4 & 3 & & 0.58 & 0.31 & 42 \\
   & 2016-06-18 & 01:02 & 36 & 17--650 & 46 & 0.5 & 1 & 4 & 3 & & 0.63 & 0.32 & 65 \\
\hline
\end{tabular}
\tablefoot{
\tablefoottext{a}{Start time of observation.}
\tablefoottext{b}{Number of ALMA 12~m antennas.}
\tablefoottext{c}{Minimum and maximum projected baseline separations.}
\tablefoottext{d}{On-source integration time.}
\tablefoottext{e}{Precipitable water vapour content of the atmosphere}
\tablefoottext{f}{Bandpass (B), amplitude (A), and phase (P) calibrators. The calibrators are: 1: J1427-4206, 2: Callisto, 3: J1058-8003, 4: J1107-4449.}
\tablefoottext{g}{Major and minor axes (FWHM) and position angle East from North of the synthesized beam at 345.8~GHz.}
}
\end{table*}

Another promising candidate has been proposed in the past. The dense core Cha-MMS1 is embedded in a filament within the Chamaeleon~I molecular cloud \citep{Belloche11} at a distance of 192~pc \citep{Dzib18}. After it has been classified as a Class 0 protostar
\citep{Reipurth96, Lehtinen01}, \citet{Lehtinen03} did not detect cm-wavelength continuum emission suggesting it to be either in an earlier protostellar phase or pre-stellar.
\citet{Belloche06} found high deuterium fractionation levels of HCO$^+$ and N$_2$H$^+$ which are
typical for evolved pre-stellar or young protostellar cores 
\citep[e.g.,][]{Crapsi05, Emprechtinger09}. In addition, the detection of Cha-MMS1 at 24~$\mu$m by the Spitzer Space Telescope led
\citet{Belloche06} to infer that a compact, hydrostatic core is already present. The comparison
of the deuterium fractionation and IR flux of Cha-MMS1 to values of various known pre-stellar cores and
Class 0 protostars led them to conclude that it might be a FHSC or an extremely young Class 0
protostar. Additionally, they failed to detect large-scale CO 3--2 outflows with the APEX telescope
suggesting that it may not have entered the Class 0 phase yet. \citet{Tsitali13} derived an
internal luminosity for Cha-MMS1 ranging from 0.08 to 0.18~$L_\odot$ by comparison with FHSC predictions
from RMHD simulations\footnote{\citet{Tsitali13} assumed a distance of $150$~pc and an inclination of Cha-MMS1 to the
line of sight of $45^\circ<i<60^\circ$. With a distance of 192~pc, this range turns into 
0.13--0.29~$L_\odot$.}. Furthermore, they found clear signatures of infall and rotation which again,
according to simulations, are required ingredients for the launch of an outflow 
\citep{Commercon12}. They also report wing emission in the central spectra of CS 2--1 and
broadened blue and red peaks of CO 3--2 which are not reproduced by their infall model. These features may
originate from a compact outflow consisting of denser, high-velocity material close to the centre of the
core but they could also be partly due to extended emission emanating from the nearby Class~I protostar
Ced~110~IRS~4 \citep{Tsitali13, Belloche06}. To find out whether Cha-MMS1 drives an
outflow on small scales and hence clarify its current evolutionary stage, we performed
observations at high angular resolution with the Atacama Large Millimetre/submillimetre Array (ALMA) and
report the results in this paper. Section~\ref{sec:obs} describes the observational setup and the data reduction
steps. The results are presented in Sect.~\ref{sec:results} and discussed in Sect.~\ref{sec:discussion}. The conclusions are given in Sect.~\ref{sec:conclusion}.

\section{Observations and data reduction} \label{sec:obs}

\subsection{Observations}

We used ALMA to observe Cha-MMS1 in CO 3--2 and $^{13}$CO 3--2 emission. The 
field was centred at ($\alpha, \delta$)$_{\rm J2000}$=
($11^{\rm h}06^{\rm m}33{\fs}13, -77^\circ23\arcmin35\farcs1$), which are the coordinates 
of the counterpart of Cha-MMS1 detected with the MIPS instrument on board of 
the \textit{Spitzer} satellite as reported by \citet{Belloche11}. The Band 7
receiver was tuned to 345.8~GHz. The size of the primary beam of the 12~m 
antennas is 16.8$\arcsec$ (HPBW) at 345.8~GHz \citep[][]{Remijan15}. The 
observations were conducted in several sessions during Cycle 3 in two 
different configurations, with baselines up to 310~m for the first 
configuration and up to 700~m for the second one (see Table~\ref{t:setup}). The shortest baselines of 15--17~m imply a 
maximum recoverable scale of about 7$\arcsec$.
The synthesized beam size at 345.8~GHz is 
$\sim$ $1.0\arcsec \times 0.8\arcsec$ for the first 
configuration and $0.6\arcsec \times 0.3\arcsec$ for the second one. The 
bandpass, amplitude, and phase calibrators are indicated in Table~\ref{t:setup}. 

The data was recorded over four spectral windows, two broadband ones for the 
continuum emission covering the frequency ranges 331.6--333.6~GHz and 
342.8--344.8~GHz with a spectral resolution of 15.6~MHz, and two narrowband ones 
covering 330.47--330.70~GHz and 345.66--345.90~GHz with a spectral resolution of 
 122.1 kHz ($\sim$0.11~km~s$^{-1}$) to target $^{13}$CO 3--2 and CO 3--2, with
rest frequencies of 330587.9653~MHz and 345795.9899~MHz, respectively.

\subsection{Data reduction and imaging}

The data was calibrated with the Common Astronomy Software Applications 
package (CASA), version 4.5.0 for the first configuration and version 4.5.3 
for the second one. We used the procedures provided by the Joint ALMA 
Observatory to apply the bandpass, amplitude, and phase calibrations. The 
splitting of the continuum and line emission was performed in the uv plane 
with the CASA task \textit{uvcontsub}, with a polynomial fit of order 1.  We 
used version 4.7.1 of CASA to image the calibrated data after merging the five 
datasets. The imaging was done with the CASA task \textit{clean} with 
\textit{Briggs} weighting, a \textit{robust} parameter of 0.5, and a threshold 
of 13 mJy for CO~3--2, 17~mJy for $^{13}$CO~3--2, and 0.15~mJy for the 
continuum. We produced cubes
and images with $512 \times 512$ pixels, a pixel size of 0.07$\arcsec$, and a channel width of 122.1 kHz.
The imaged data has an rms of 3.6~mJy/beam and 4.7~mJy/beam per channel for CO~3--2 and $^{13}$CO~3--2, respectively, and 0.053~mJy/beam for the continuum emission. The 
synthesized beams (HPBW) are $0.58\arcsec \times 0.37\arcsec$, 
$0.62\arcsec \times 0.41\arcsec$, and $0.59\arcsec \times 0.38\arcsec$ with 
position angles East from North of 47$^{\circ}$, 48$^{\circ}$, and 47$^{\circ}$, 
respectively.
During the course of data analysis in the subsequent sections, we used the data cubes with different spectral resolutions. The smoothing was done using the \textit{smooth box} command of the GILDAS/CLASS software which computes the average intensity of a number of channels.

\section{Results}\label{sec:results}

\begin{figure*}
    \centering
    \includegraphics[width=0.95\textwidth]{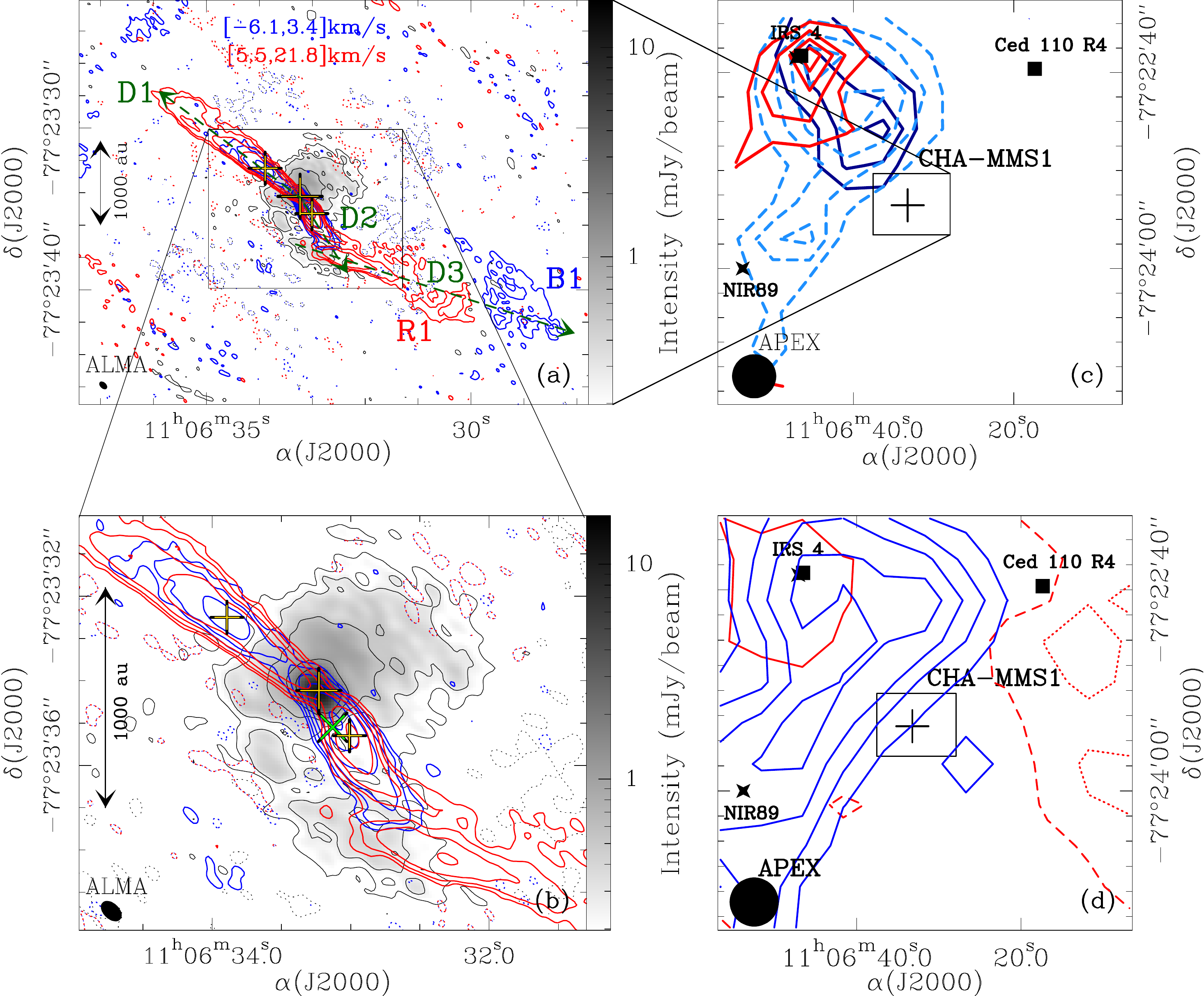}
    \caption{\textbf{a)} CO 3--2 integrated intensities (blue and red contours) observed with ALMA toward Cha-MMS1. The continuum emission is shown in greyscale with black contours. The large yellow cross marks its peak position. The CO emission is integrated over the velocity ranges $[-6.1,3.4]$~km~s$^{-1}$ (blue) and $[5.5,21.8]$~km~s$^{-1}$ (red). The contours are $-2.5\sigma$, $2.5\sigma$, $5\sigma$, and then increase by a factor of two at each step with $\sigma=7.04$~mJy/beam~km/s, $9.06$~mJy/beam~km/s, and $0.05$~mJy/beam for the blueshifted, redshifted, and the continuum emission, respectively. The HPBW is shown in the bottom left corner. The yellow crosses mark the positions of the spectra shown in Fig.~\ref{fig:spectra}. The green dashed arrows indicate the axes along which position-velocity diagrams were taken for Figs.~\ref{fig:PV} and \ref{fig:PV_ext}. B1 and R1 term the less-collimated blue- and redshifted structures, respectively. \textbf{b)} Same as a) but zoomed in on the central region. The phase centre is marked with a green cross in this panel. \textbf{c)} CO 3--2 integrated intensities observed with APEX toward Cha-MMS1 (cross) \citep[adapted from][]{Belloche06}. The redshifted [6.6,9.8]~km~s$^{-1}$, blueshifted [-0.2,1.0]~km~s$^{-1}$, and less blueshifted [1.2,2.4]~km~s$^{-1}$ emission are plotted as solid red, solid blue, and dashed blue contours, respectively. The first contour and step are $3\sigma$, $6\sigma$, and $3\sigma$ with $\sigma=0.13, 0.13$, and $0.24$~K~km/s, respectively, as in \citet{Belloche06}. The HPBW is shown in the bottom left corner. The star and square symbols mark positions of NIR \citep{Persi01} and $3.5$~cm sources \citep{Lehtinen03}, respectively. \textbf{d)} Same as c) but for CO 3--2 integrated intensities at velocities close to $\varv_\mathrm{sys}$. The redshifted [5.5,6.6]~km~s$^{-1}$ emission is shown as dotted ($3\sigma$), dashed ($6\sigma$), and solid ($9\sigma$) contours with $\sigma=0.13$~K~km~s$^{-1}$. The blueshifted [2.4,3.4]~km~s$^{-1}$ emission is plotted as solid contours. The first contour and the steps are $6\sigma$ with $\sigma=0.13$~K~km~s$^{-1}$. The small box corresponds to the field of view shown in a).}
    \label{fig:co_int_map}
\end{figure*}

\subsection{Continuum emission}
\label{ss:continuum}

The map of continuum emission averaged over both sidebands is shown in 
Fig.~\ref{fig:co_int_map} in greyscale. The emission is centrally peaked. We performed
an elliptical Gaussian fit in the uv plane in order to derive the peak 
position, size, and flux density of the compact emission. We used the task 
uv\_fit of the GILDAS/MAPPING 
software\footnote{see http://www.iram.fr/IRAMFR/GILDAS.} for this purpose. The 
fit was performed on the visibilities of each sideband separately. We obtain 
an average peak position of $11^{\rm h}06^{\rm m}33\fs231 -77^\circ23\arcmin34\farcs23$ 
and an average emission size (FWHM) of
$(0.61\arcsec \pm 0.01\arcsec) \times (0.57\arcsec \pm 0.02\arcsec)$
with a position angle $50^\circ \pm 10^\circ$. 
At a distance of 192~pc, the geometrical mean of the angular sizes translates 
into a radius of $\sim$55~au. The total flux density of the 
fitted Gaussian is $24.0 \pm 0.1$~mJy and $25.1 \pm 0.1$~mJy in the lower and 
upper sidebands, respectively. The cleaned map of the residual visibilities 
contains a weak ($< 1$~mJy/beam), extended ($\sim$0.5$\arcsec$) structure, a 
weak ($> -1$~mJy/beam), negative residual ring, and an unresolved component 
at the peak position, with a peak flux density of $\sim$2~mJy/beam.
The residual negative ring and unresolved component may both result from our assumption of a Gaussian shape to fit the envelope in the uv plane while it may actually have a power-law structure.

Following Appendix~B.4 of \citet{Belloche11}, we assume a dust mass 
opacity $\kappa_{\rm 345~GHz} = 0.02$~cm$^2$~g$^{-1}$ of gas to convert the total flux 
density of the fitted Gaussian component into a gas mass. With a luminosity of 
$\sim$0.2~$L_\odot$, we expect a dust temperature on the order of 30~K at a 
radius of 55~au \citep[][]{Terebey93,Motte01}. With these assumed dust mass 
opacity and temperature, we obtain a mass of $2.7 \times 10^{-3}$~$M_\odot$ 
provided the continuum emission is optically thin. The flux density of
the Gaussian component derived above corresponds to about 1~K in temperature 
scale. This is much smaller than the assumed dust temperature and thus 
confirms that the dust emission is still optically thin at the scales traced
here with ALMA.

\begin{figure}
    \centering
    \includegraphics[scale=0.3]{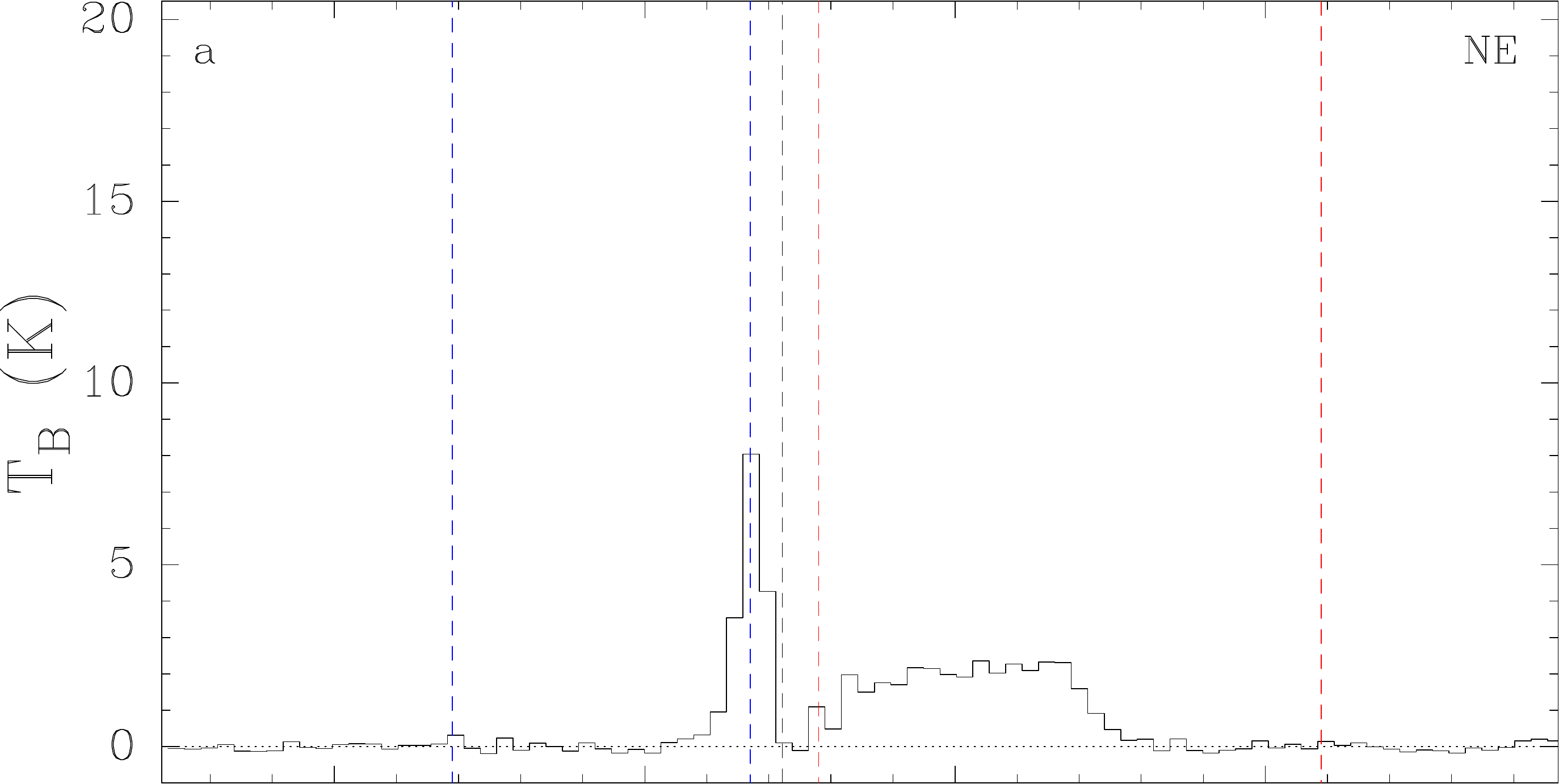}\\[0.1cm]
     \includegraphics[scale=0.3]{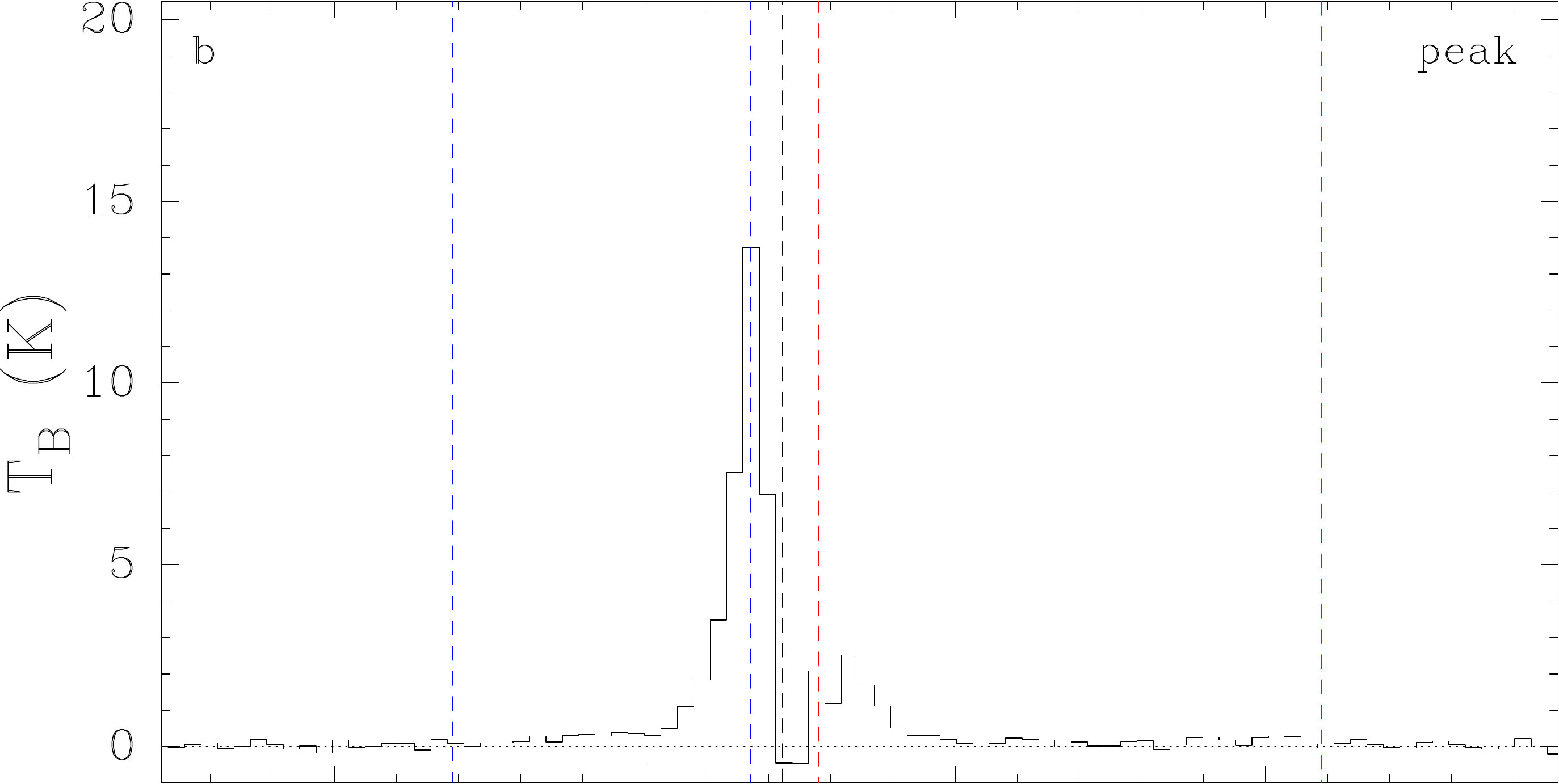}\\[0.1cm]
      \includegraphics[scale=0.3]{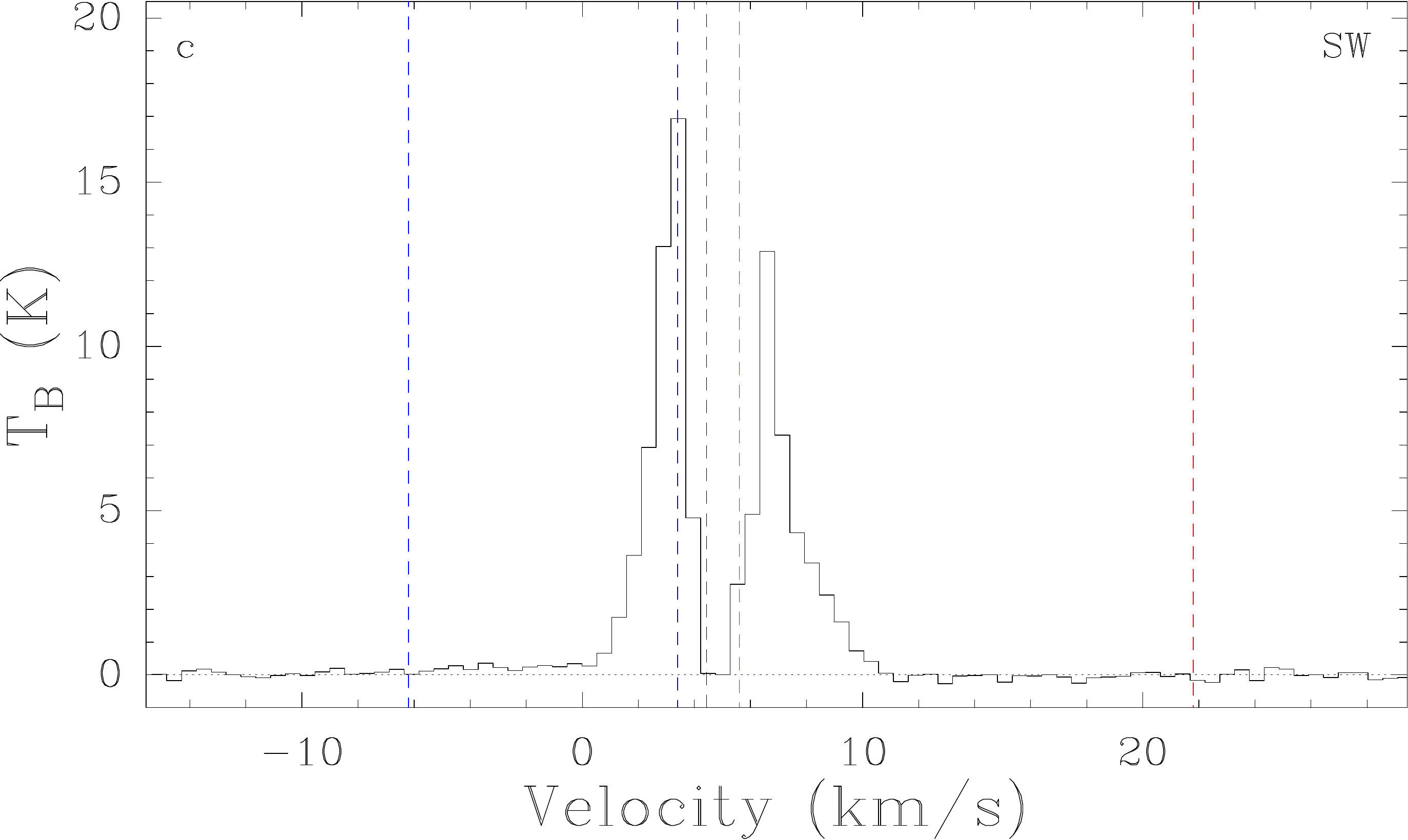}
    \caption{CO 3--2 spectra observed with ALMA toward Cha-MMS1 in brightness temperature scale and smoothed over five channels. The spectra are taken toward three positions marked with yellow crosses in Fig.~\ref{fig:co_int_map}: a) in the NE lobe, b) at the peak of the continuum emission, and c) in the SW lobe. The red and blue dashed lines correspond to the integration intervals defined in Sect.~\ref{sec:morphology} while the black dashed line indicates the systemic velocity of Cha-MMS1, $\varv_\mathrm{sys}=4.43$~km~s$^{-1}$ \citep{Belloche06}.}
    \label{fig:spectra}
\end{figure}
\begin{figure}
    \includegraphics[width=0.48\textwidth]{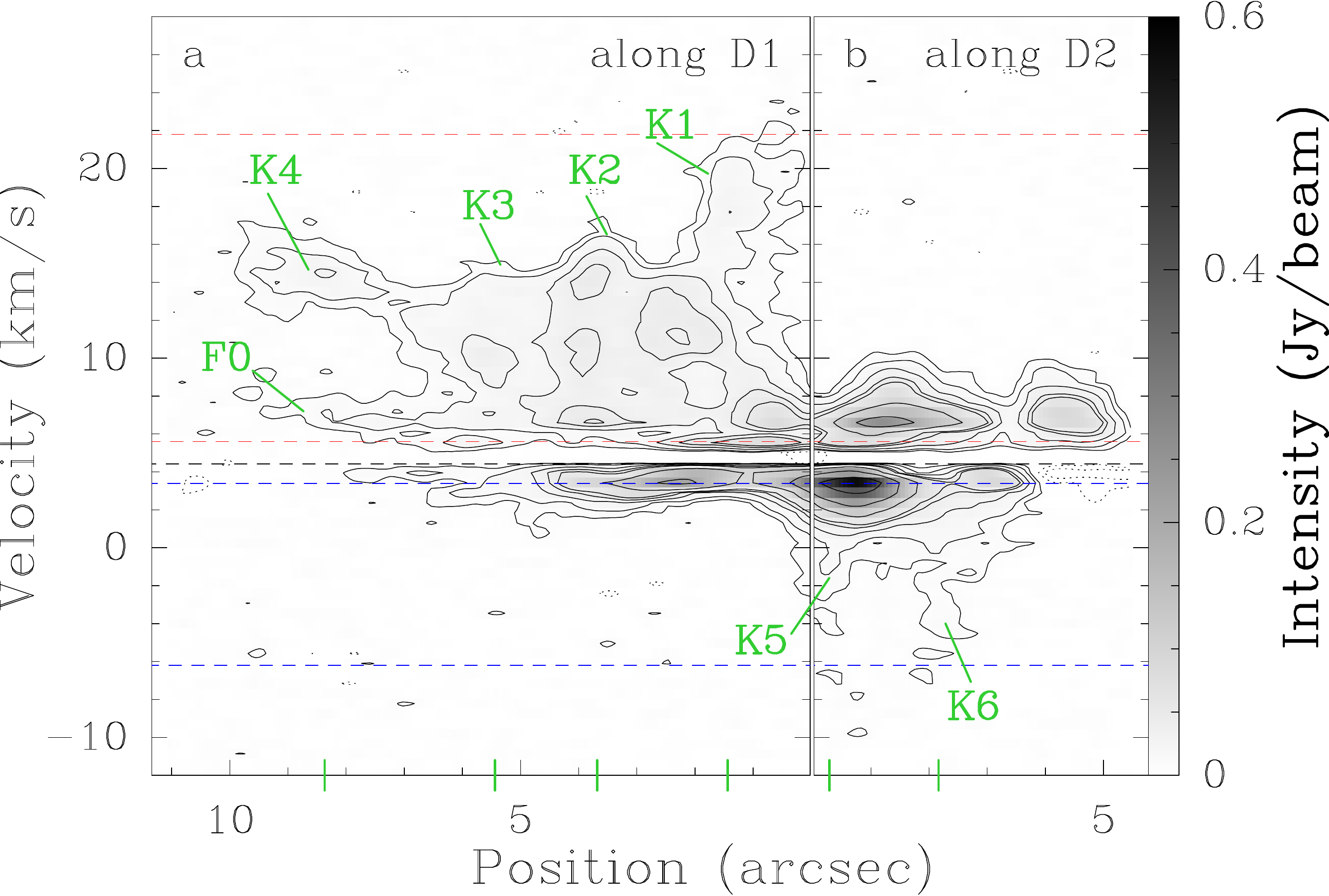}\\[-0.5cm]  
      
    \hspace{-0.7cm}\includegraphics[width=.55\textwidth]{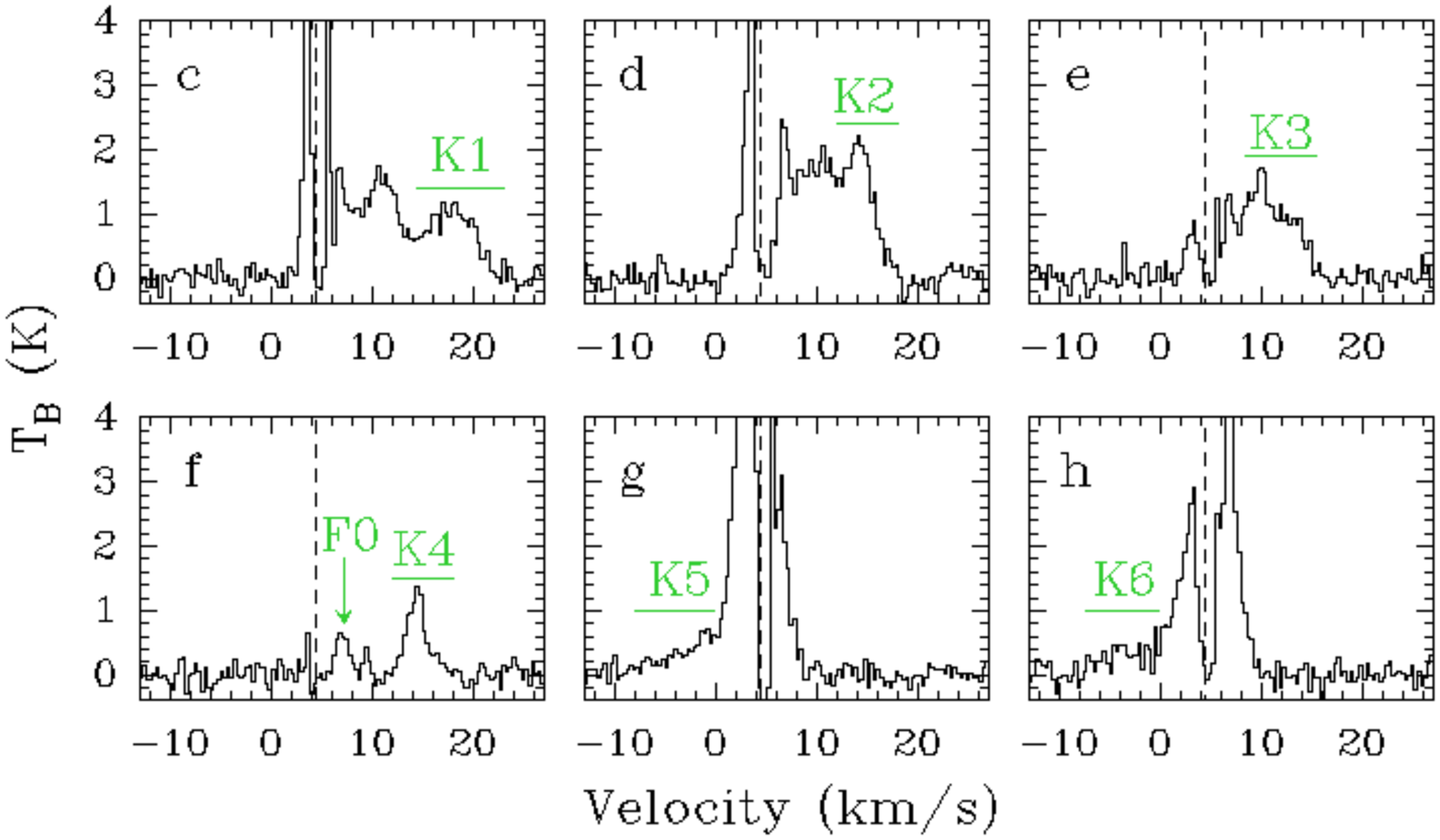}\\[-1.2cm]
    \caption{\textbf{a)-b)} CO 3--2 position-velocity diagrams along the outflow axes a) D1 and b) D2 shown as green dashed arrows in Fig.~\ref{fig:co_int_map}a. The origin of the position axes corresponds to the continuum peak. The intensity is smoothed over five channels. The contours are $-6\sigma$, $-3\sigma$, $3\sigma$, $6\sigma$, $12\sigma$, $18\sigma$, $24\sigma$ $48\sigma$, $96\sigma$, and $192\sigma$ with $\sigma=2.0$~mJy/beam. The red and blue dashed lines show the inner and outer integration limits. \textbf{c)-h)} CO 3--2 spectra in brightness temperature scale taken along D1 and D2 at the positions marked as green ticks on the x-axis in the PV diagrams in a) and b). K1--K6 denote the bumps corresponding to the bullets and F0 the bulk flow. The intensity is smoothed over two channels. In all panels, the black dashed line indicates the systemic velocity of Cha-MMS1.}
    \label{fig:PV}
\end{figure}

\subsection{Detection of a fast outflow}\label{sec:morphology}

Figure~\ref{fig:spectra}b shows the CO 3--2 spectrum toward the position of the continuum peak
determined in Sect.~\ref{ss:continuum} and marked as a large yellow cross in Fig.~\ref{fig:co_int_map}a. It is located at (0.33$\arcsec$,0.87$\arcsec$) from the phase centre.
The lines are broad with strong wing emission suggesting outflow-like structures to be present.
Figure~\ref{fig:co_int_map}a shows CO 3--2 integrated intensity maps which reveal a bipolar outflow originating from the centre of Cha-MMS1 with an axis approximately along the northeast (NE) southwest
(SW) direction. Figure~\ref{fig:co_int_map}b zooms in on the continuum and outflow emission to better display the structures close to the centre. The inner integration limits used in the integrated intensity maps in Fig.~\ref{fig:co_int_map}a and \ref{fig:co_int_map}b are chosen such that the outflow emission
remains clearly distinguishable from the emission of the surrounding envelope in the channel maps, with values of 3.4~km~s$^{-1}$ and 5.5~km~s$^{-1}$ for the blue- and redshifted emission, respectively. The outer integration limits are determined by investigating the CO data cube containing spectra smoothed over
eight channels (after imaging). The smoothed channel maps are shown in Fig.~\ref{Appfig}. We search for faint emission
which can still be associated with the outflow. We find a minimum velocity  of 
$-$6.1~km~s$^{-1}$ (blue) at $(-1.2\arcsec,-1.8\arcsec)$ from the continuum peak and a maximum velocity of
21.8~km~s$^{-1}$ (red) at $(0.5\arcsec,0.4\arcsec)$ from the continuum peak which we set as the outer
integration limits. This yields velocities of $-$10.6~km~s$^{-1}$ and 17.4~km~s$^{-1}$ with respect to the systemic velocity $\varv_\mathrm{sys}= 4.43$~km~s$^{-1}$ \citep{Belloche06} for the blue- and redshifted emission, respectively. 

In Figs.~\ref{fig:spectra}a and \ref{fig:spectra}c, we show additional CO 3--2 spectra toward positions of peak outflow intensities marked as small yellow crosses in Figs.~\ref{fig:co_int_map}a and \ref{fig:co_int_map}b. The spectrum in 
Fig.~\ref{fig:spectra}a is located at $(1.8\arcsec,1.6\arcsec)$ from the continuum peak position in the NE
lobe. It shows lower peak intensities compared to the other two spectra but reveals significant emission at high velocities. The spectrum 
in Fig.~\ref{fig:spectra}c is located at $(-0.7\arcsec,-1.1\arcsec)$ from the continuum peak position in the SW lobe. It shows higher intensities compared to the other spectra and even broader wings.

Figures~\ref{fig:PV}a and \ref{fig:PV}b show the position-velocity (PV) diagrams along the outflow axes indicated as green dashed arrows and labelled D1 and D2 in Fig.~\ref{fig:co_int_map}a. The morphology of the outflow and PV diagrams are similar to case
3 of \citet{Cabrit90} where both lobes contain red- as well as blueshifted emission indicating that it
is oriented almost in the plane of the sky. We give an estimate of the inclination in Sect.~\ref{sec:props}.

The PV diagrams reveal that the outflow emission covers a wide range of velocities, especially the redshifted emission along D1 and it shows bumps toward higher velocities which we label K1-K6. These bumps suggest the occurrence of episodic ejections. In Figs.~\ref{fig:PV}c-h, we show spectra taken along D1 and D2 indicated as green ticks on the x-axis in the PV diagrams. 
These spectra reveal multiple peaks that are characteristic of bullets resulting from an intermittent ejection process \citep[see][]{Bachiller96}. We interpret the contour at $\sim$8~km~s$^{-1}$ marked as F0 in Fig.~\ref{fig:PV}a as tracing the slow-moving cavity at the interface with the ambient medium inside which the fast bullets propagate.

Assuming that the high velocities represent those of bullets rather than the bulk velocity of the outflow, we define maximum velocities such that the full length of each outflow component is still maintained using the channel-maps in Fig.~\ref{Appfig}. We will use these values in Sect.~\ref{sec:vmax}. We find $\varv^\prime_\mathrm{max}=\vert \varv_\mathrm{max}-\varv_\mathrm{sys} \vert = 2.6$~km~s$^{-1}$ and 2.1~km~s$^{-1}$ for the blueshifted emission in the
SW and NE lobes, respectively, and 4.7~km~s$^{-1}$ and 
11.5~km~s$^{-1}$ for the redshifted emission in the SW and NE lobes, respectively. However, as mentioned above (F0), the bulk velocity of the redshifted emission in the NE lobe may rather be $\varv^\prime_\mathrm{max}=3.6$~km~s$^{-1}$ with respect to the systemic velocity. 

Further, we measure projected spatial extents for each lobe which range from 4$\arcsec$ to
13$\arcsec$ where we did not include the separated emission labelled B1 in Fig.~\ref{fig:co_int_map}a. These values yield projected outflow lengths of 800--2500~au at a distance of 192~pc.
The observed velocities and the length of each lobe are summarised in Table~\ref{tab:props}.

\subsection{Morphology of the outflow}

The maps in Figs.~\ref{fig:co_int_map}a and \ref{fig:co_int_map}b further show that the NE and SW lobe possess different position angles. The axis of the NE lobe is at $\sim$50$^\circ$ while the axis of the SW lobe is at $\sim-$150$^\circ$ which means that they are tilted by $\sim$20$^\circ$ with respect to each other. Regarding the redshifted emission, the axis of the SW lobe reveals another conspicuous change of position angle to $\sim-$120$^\circ$ at ($-1.9\arcsec,-1.9\arcsec$) from the continuum peak position. The emission along this second axis in the SW lobe (labelled R1 in Fig.~\ref{fig:co_int_map}a) gets less and less collimated with distance to the centre. Farther in the southwest, there is also extended blueshifted emission labelled B1 in Fig.~\ref{fig:co_int_map}a which appears at low velocities (see also Fig.~\ref{Appfig}). Figure~\ref{fig:PV_ext} shows the PV diagram along R1 and B1 which is taken along the green dashed arrow labelled D3 in Fig.~\ref{fig:co_int_map}a. The arrow covers exactly the range of positions displayed in the PV-diagram. In Sect.~\ref{Sect:extended}, we discuss these peculiarities in the morphology in more detail.

Figure~\ref{fig:co_int_map}c shows CO 3--2 integrated intensity maps obtained on larger scales with the single-dish APEX
telescope reported by \citet{Belloche06}. They are dominated by an outflow driven by the
nearby Class~I protostar IRS~4. Additionally, in Fig.~\ref{fig:co_int_map}d, we show the APEX emission integrated over [5.5,6.6]~km~s$^{-1}$ and
[2.4,3.4]~km~s$^{-1}$ to include the lowest velocities comprised in the ALMA integration ranges. This emission reveals large-scale structures probably tracing the ambient material in the filament which is filtered out in the interferometric maps. 
\begin{table}
\caption{Properties of each lobe of the outflow of Cha-MMS1 determined from CO 3--2 ALMA observations.\label{tab:props} }             
\centering      
\begin{tabular}{l l c c c c } 
\hline\hline \\[-0.3cm]
\multicolumn{2}{c}{Lobe} &  $\varv_\mathrm{max}$\tablefootmark{a} & $\varv_\mathrm{max}'$\tablefootmark{b} & $R^\mathrm{proj}_\mathrm{lobe}$\tablefootmark{c} & $R^\mathrm{proj}_\mathrm{lobe}$\tablefootmark{d} \\[0.1cm]
& & (km~s$^{-1}$) & (km~s$^{-1}$) & ($\arcsec$) & ($10^{3}$~au) \\[0.1cm]
\hline \\[-0.3cm]
\multirow{3}{*}{Red} & NE & 15.9 & 11.5 & 12.5 & 2.4 \\
& NE-F0\tablefootmark{*} & 8.0 & 3.6 & 12.5 & 2.4 \\
& SW & 9.1 & 4.7 & 13.0 & 2.5 \\
\multirow{2}{*}{Blue} & NE & 2.3 & 2.1 & 7.4 & 1.4 \\
& SW & 1.8 & 2.6 & 4.2 & 0.8 \\
\hline
\end{tabular}
\tablefoot{\tablefoottext{a}{Minimum (blueshifted) and maximum (redshifted) velocities for which the full length of a lobe is still maintained.} \tablefoottext{b}{Maximum velocities $\varv^\prime_\mathrm{max}=\vert \varv_\mathrm{max}-\varv_\mathrm{sys}\vert $ with $\varv_\mathrm{sys}=4.43$~km~s$^{-1}$.}
\tablefoottext{c}{Projected length of a lobe along the outflow axis.} \tablefoottext{d}{Projected and deconvolved (beam major axis 0.58$\arcsec$) length of a lobe at a distance of $D=192$~pc.}\\
\tablefoottext{*}{NE-F0 represents the case where the bulk velocity of the NE lobe is traced by F0 in Fig.~\ref{fig:PV}a.}}
\end{table}

\subsection{Inclination of the outflow}\label{sec:props}

Based on the integrated intensity maps and the PV diagrams we already concluded that the outflow axis is almost perpendicular to the line of sight. 
The inclination $i$ is in the range $90^\circ-\nicefrac{\theta}{2} < i < 90^\circ+\nicefrac{\theta}{2}$ where $\theta$ is the opening angle of the outflow. 
By using geometry, we determine $\theta = 22^\circ \pm 3^\circ$ for the more extended northeastern redshifted lobe which we adopt for both lobes. 
Hence, the inclination angle of Cha-MMS1 is in the range $79^\circ < i < 101^\circ$. 

We use the minimum value of 79$^\circ$  to deproject the velocities. Considering the highest velocities reported in Sect.~\ref{sec:morphology}, we find maximum deprojected velocities of 55.2~km~s$^{-1}$ for the blueshifted emission and 91.0~km~s$^{-1}$ for the redshifted emission. If the true
inclination is higher than 79$^\circ$, the velocities will be even higher.

\subsection{Optical depth}\label{sec:depth}

\begin{figure}
    \centering
    \includegraphics[width=0.49\textwidth]{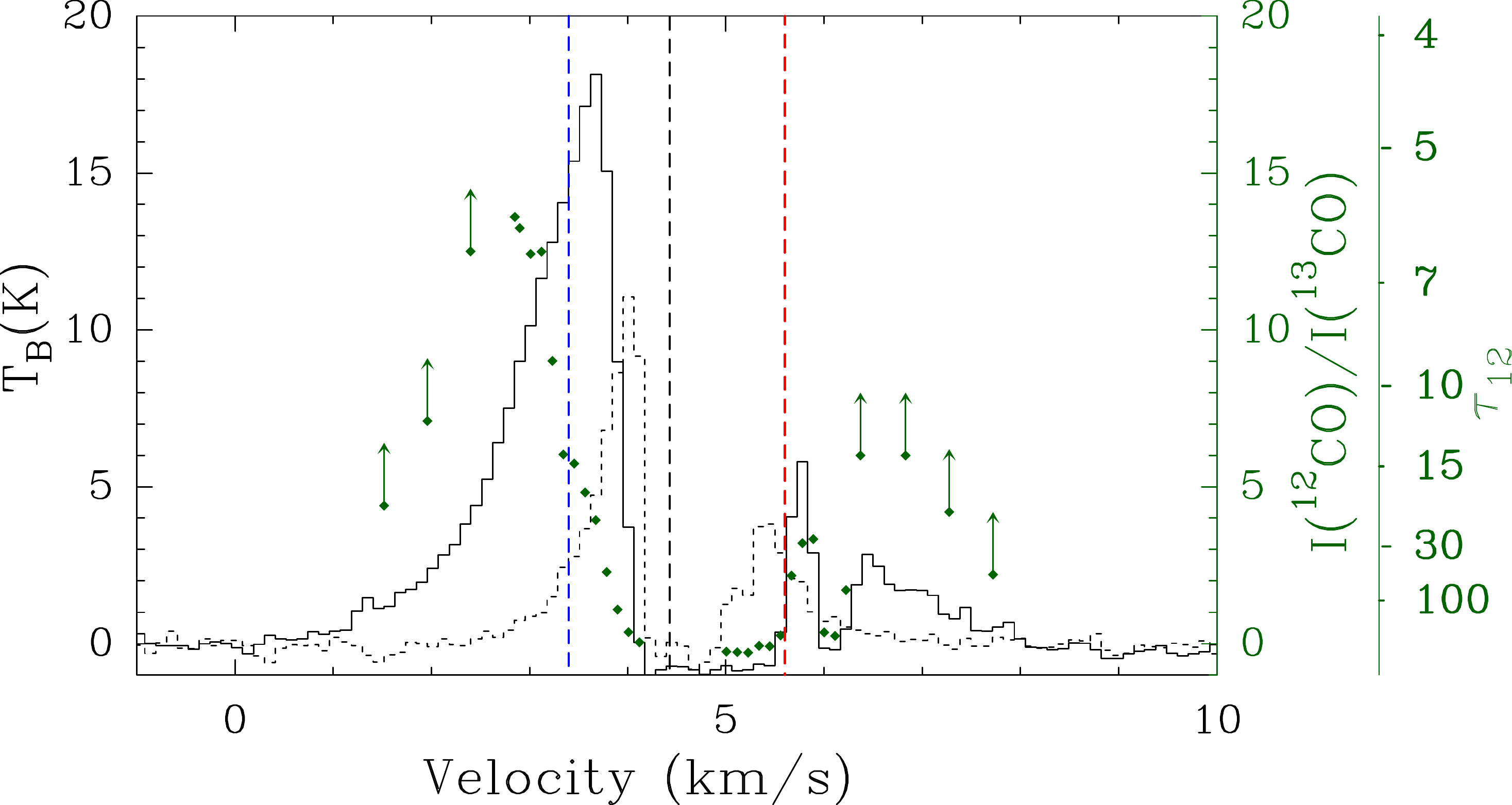}
    \caption{CO 3--2 (solid black) and $^{13}$CO 3--2 (dotted black) spectra at the peak position of Cha-MMS1 in brightness temperature scale. Green square symbols show the ratio of CO to $^{13}$CO intensity while green arrows indicate lower limits of this ratio after smoothing over four channels. The scale is given on the first right axis. The optical depth scale is shown on the second right axis. The red and blue dashed lines mark the inner integration limits while the black dashed line indicates the systemic velocity of Cha-MMS1.}
    \label{fig:optical_depth}
\end{figure}

\begin{table*}[h!]
\caption{Mass and momentum of the outflow.}
\label{tab:M_CO}
\centering
\begin{tabular}{l c c c c c c c c} 
\hline\hline\\[-0.3cm] 
Lobe  & \multicolumn{2}{c}{$M_\mathrm{obs}$\tablefootmark{a}} & \multicolumn{2}{c}{$M_\mathrm{corr}$\tablefootmark{b}} & \multicolumn{2}{c}{$P_\mathrm{obs}$\tablefootmark{c}} & \multicolumn{2}{c}{$P_\mathrm{corr}$\tablefootmark{d}} \\
 & \multicolumn{2}{c}{($10^{-4}~M_\odot$)} & \multicolumn{2}{c}{($10^{-4}~M_\odot$)} & \multicolumn{2}{c}{($10^{-4}~M_\odot$~km~s$^{-1}$)} & \multicolumn{2}{c}{($10^{-4}~M_\odot$~km~s$^{-1}$)} \\
 
& $\tau_\mathrm{12}<<1$ & $\tau_\mathrm{12}=5$ & $<<1$ & $5$ & $<<1$ & $5$ & $<<1$ & 5 \\
\hline\\[-0.3cm]
NE & 0.3 & 1.0 & 1.2 & 4.6 & 1.1 & 4.9 & 5.1 & 23.9 \\
SW & 1.0 & 1.7 & 4.4 & 7.5 & 1.2 & 3.0 & 5.8 & 14.2 \\
\hline
\end{tabular}
\tablefoot{\tablefoottext{a}{Observed outflow mass computed from $^{13}$CO where detected, and $^{12}$CO elsewhere assuming either optically thin ($\tau_\mathrm{12}<<1$) or thick ($\tau_\mathrm{12}=5$) CO emission.} \tablefoottext{b}{Outflow mass corrected for missing material at ambient velocities and hidden atomic material.} \tablefoottext{c}{Observed outflow momentum computed with the same approach as for the mass.} \tablefoottext{d}{Outflow momentum corrected for missing material at ambient velocities and hidden atomic material.}}
\end{table*}

\begin{table*}[h!]
\caption{Dynamical time and momentum of the outflow from the perpendicular method.}
\label{tab:F_CO}
\centering
\begin{tabular}{l c c c c c c c c c c c c c c c c c} 
\hline\hline\\[-0.3cm] 
 & $W_\mathrm{lobe}$\tablefootmark{a} & \multicolumn{2}{c}{$\left\langle \varv^\prime \right\rangle$\tablefootmark{b}} & \multicolumn{2}{c}{$t_\mathrm{perp}$\tablefootmark{c}} &  \multicolumn{4}{c}{$t_\mathrm{corr}$\tablefootmark{d}} & \multicolumn{2}{c}{$F_\mathrm{perp}$\tablefootmark{e}} & \multicolumn{4}{c}{$F_\mathrm{corr}$\tablefootmark{f}}\\
 & ($10^2$~au) & \multicolumn{2}{c}{(km~s$^{-1}$)} & \multicolumn{2}{c}{($10^2$~yr)} & \multicolumn{4}{c}{($10^2$~yr)} & \multicolumn{2}{c}{($10^{-6}~M_\odot$~km~s$^{-1}$~yr$^{-1}$)} & \multicolumn{4}{c}{($10^{-6}~M_\odot$~km~s$^{-1}$~yr$^{-1}$)} \\
 $\tau_\mathrm{12}$ & & $<<1$ & $5$ & $<<1$ & $5$ & \multicolumn{2}{c}{$<<1$} & \multicolumn{2}{c}{$5$} & $<<1$ & $5$ & \multicolumn{2}{c}{$<<1$} & \multicolumn{2}{c}{$5$} \\
 $i$ & & & & & & 60$^\circ$ & 90$^\circ$ & 60$^\circ$ & 90$^\circ$ & & &  60$^\circ$ & 90$^\circ$ & 60$^\circ$ & 90$^\circ$ \\
\hline\\[-0.3cm]
NE & 3.3 & 3.8 & 4.9 & 1.4 & 1.1 & 2.1 & 1.6 & 1.7 & 1.3 & 0.8 & 4.6 & 2.0 & 3.7 & 11.4 & 21.6 \\
SW & 2.0 & 1.2 & 1.7 & 2.6 & 1.8 & 4.1 & 3.1 & 2.8 & 2.2 & 0.5 & 1.6 & 1.2 & 2.2 & 4.1 & 7.7 \\
\hline
\end{tabular}
\tablefoot{\tablefoottext{a}{Maximum deconvolved (beam minor axis $0.37\arcsec$) radius of a lobe.} \tablefoottext{b}{Intensity-weighted velocity for both opacity assumptions.} \tablefoottext{c}{Dynamical age calculated with the perpendicular method (DC07) for both opacity assumptions.} \tablefoottext{d}{$t_\mathrm{perp}$ corrected for projection effects.} \tablefoottext{e}{Momentum force calculated with the perpendicular method for both opacity assumptions.}\tablefoottext{f}{$F_\mathrm{perp}$ corrected for projection effects and hidden atomic material.}}
\end{table*}

Both, CO 3--2 as well as $^{13}$CO 3--2 show strong self-absorption close to $\varv_\mathrm{sys}$
indicating high optical depths (see Figs.~\ref{fig:spectra} and \ref{fig:optical_depth}). To find out whether the CO emission remains optically thick in the wings
of the spectral line, that is in the outflow, we compute the ratio of CO to $^{13}$CO emission. Figure \ref{fig:optical_depth} shows the spectra of CO and $^{13}$CO toward the continuum peak position with the ratio of
the intensities overlaid as green square symbols. We plot the ratio only for values of $I(^{13}\mathrm{CO})$ higher than 2.5$\sigma$. Unfortunately, the sensitivity of our $^{13}$CO data is not high
enough to provide significant ratios far out in the wings. Therefore, we average the $^{13}$CO wing
emission over four channels and use the smoothed spectra down to 3$\sigma$. For channels with an intensity below this threshold, we use 3$\sigma$ as an upper limit to $I(^{13}\mathrm{CO})$. We pursue this calculation until $I(\mathrm{CO})\approx5\sigma$. The resulting
lower limits of the ratio are indicated with green arrows in Fig.~\ref{fig:optical_depth}. Based on the
intensity ratio, the optical depth $\tau_\mathrm{12}$ can be estimated using
\begin{equation}
	\frac{I(\mathrm{CO})}{I(^{13}\mathrm{CO})} = \frac{1-\mathrm{e}^{-\tau_{12}}}{1-\mathrm{e}^{-\tau_{13}}} \label{eq:1}
\end{equation}
with $\tau_{13}=\tau_{12}/X$ where $X$ is the local ISM $^{12}$CO/$^{13}$CO isotopic ratio equal to 77 \citep{Wilson94}. The opacity scale is displayed on the outer
right axis of Fig.~\ref{fig:optical_depth}. Toward the continuum peak position, we find high optical depths $\tau_{12} \gtrsim 100$ for the emission
close to $\varv_\mathrm{sys}$ as expected. At the inner integration limit, the optical depth is still at $\sim$15 for the blueshifted emission and it decreases to 6 at $\sim$2.8~km~s$^{-1}$. The optical depth of the redshifted emission remains higher than 30 up to at least $\sim$6.2~km~s$^{-1}$. The lack of sensitivity of the $^{13}$CO observations prevents us to derive the opacity below 2.8~km~s$^{-1}$ in the blue wing and above 6.2~km~s$^{-1}$ in the red wing. 
Hence, for the following calculations, we will use the $^{13}$CO as well as the  $^{12}$CO emission (see \citet{Zhang16} for a similar strategy applied to the HH~46/47 outflow). For each pixel in the outflow, the $^{13}$CO emission is smoothed over four channels, corrected for optical depth where possible using Eq.~\ref{eq:1}, and integrated until the intensity reaches a certain threshold. When the emission drops below this threshold, we use the $^{12}$CO emission where we assume two cases,
an average optical depth of $\tau_{12} = 5$ and optically thin emission.

\subsection{Outflow mass}\label{ss:mass}

We derive the outflow mass using integrated intensities which are first converted to column densities and subsequently, to mass by using the following equation:
\begin{equation}\label{eq:mass}
M_\mathrm{obs} = K\sum_{\mathrm{lobe}}\left(X\int_{\varv_\mathrm{in}}^{\varv_\mathrm{th}}F^{13}_\tau I(^{13}\mathrm{CO})~\mathrm{d}\varv+\int_{\varv_\mathrm{th}}^{\varv_\mathrm{out}}F^{12}_\tau I(^{12}\mathrm{CO})~\mathrm{d}\varv\right),
\end{equation}
where the sum is over all pixels in the respective lobe (see below and Fig.~\ref{Appfig}), $F^{12}_\tau$ and $F^{13}_\tau=\tau_{13}/(1-\mathrm e^{-\tau_{13}})$ are correction factors for optical depth, and $K=5.07\times10^{-10}~M_\odot$~s~K$^{-1}$~km$^{-1}$ is the conversion factor from intensity to mass. The calculation of $K$ is adopted from \citet{vdM13} (hereafter vdM13) where we assume an excitation temperature of 30~K and a CO abundance of $8.3\times 10^{-5}$ relative to H$_2$ (vdM13). 
As mentioned in Sect.~\ref{sec:depth}, we account for the high optical depths at the
inner velocity cut-offs by using the $^{13}$CO emission for the mass calculation. Alike the inner integration limits for CO, we define them for $^{13}$CO by investigating the channel maps which we show in Fig.~\ref{Appfig2}. We detect $^{13}$CO outflow emission down to $\varv_\mathrm{in}=4.9$~km~s$^{-1}$ and 3.8~km~s$^{-1}$ for the red- and blueshifted emission, respectively.
Starting from these velocities, we integrate the $^{13}$CO emission as long as it is detected above 3.4$\sigma$ and 3.8$\sigma$ for the blue- and redshifted emission, respectively, and multiply it by the isotopic ratio $X=77$. The thresholds are chosen such that we certainly avoid the contribution of noise to the calculation. 
Since the $^{13}$CO emission also shows high optical depths close to the systemic velocity, we apply the correction factor $F^{13}_\tau$ whenever the intensity ratio of the isotopologues drops below 3.9, that is whenever the correction becomes significant.
In some channels the ratio drops below 1 (see, e.g., Fig.~\ref{fig:optical_depth} at $\sim$6~km~s$^{-1}$). In these cases we do not apply any correction factor and assume optically thin emission.
When the $^{13}$CO emission drops below the intensity thresholds at $\varv_\mathrm{th}$, we use the CO emission and integrate to the outer integration limits $\varv_\mathrm{out}$ defined in Sect.~\ref{sec:morphology}. Here, we apply the correction $F^{12}_\tau$ assuming the two cases of either optically thin ($\tau_\mathrm{12}<<1$) or thick emission ($\tau_\mathrm{12}=5$). The calculation is applied to every position in the respective lobe. We define masks for the red- and blueshifted emission of the NE and SW lobes, respectively, based on the integrated intensity maps in Fig.~\ref{fig:co_int_map}a. The used masks are shown in the first panel in Fig.~\ref{Appfig}.
The resulting observed outflow masses are summarised in Table~\ref{tab:M_CO} and lie in ranges of $M_\mathrm{obs}=(0.3-1.0)\times 10^{-4}~M_\odot$ and $(1.0-1.7)\times 10^{-4}~M_\odot$ for the NE and SW lobes, respectively.

In order to get the true mass, we apply two correction factors introduced by \citet{Downes07} (hereafter DC07). First, we want to account for material at low velocities which we may miss because its velocity is too close to the systemic velocity of Cha-MMS1. This correction factor depends on two parameters, the ratio of the jet to the ambient material density $\eta$ and inclination. At the scale where we detect the outflow of Cha-MMS1, we assume the outflow to be less dense than the ambient material, with $\eta=0.1$. The inclination estimate of Cha-MMS1 is in between $\alpha=0^\circ$ and 30$^\circ$ with $\alpha=90-i$, hence, we use an average value of 0.35$^{-1}$ (see column 8 in Table~3 in DC07). The second correction factor 0.64$^{-1}$ (column 2 of Table~1 in DC07) accounts for hidden atomic material. These corrections increase the outflow masses to $M_\mathrm{corr}=(1.2-4.6)\times 10^{-4}~M_\odot$ and $(4.4-7.5)\times 10^{-4}~M_\odot$ for the NE and SW lobes, respectively. The values are listed in Table~\ref{tab:M_CO}.

\subsection{Outflow momentum}

Similarly to the calculation of the outflow mass, we also determine the momentum of the outflow by  
\begin{equation}\label{eq:mom}
P_\mathrm{obs} = K\sum_{\mathrm{lobe}}\left(X\int_{\varv_\mathrm{in}}^{\varv_\mathrm{th}} F^{13}_\tau I(^{13}\mathrm{CO})~\varv^\prime~\mathrm{d}\varv+\int_{\varv_\mathrm{th}}^{\varv_\mathrm{out}} F^{12}_\tau I(^{12}\mathrm{CO})~\varv^\prime~\mathrm{d}\varv\right)
\end{equation}
where before integrating, we multiply the intensity of each channel by the corresponding velocity $\varv^\prime=\vert\varv-\varv_\mathrm{sys}\vert$. Depending again on optical depth of the CO emission, the observed momentum lies in a range of $P_\mathrm{obs}=(1.1-4.9)\times 10^{-4}~M_\odot$~km~s$^{-1}$ and $(1.2-3.0)\times 10^{-4}~M_\odot$~km~s$^{-1}$ for the NE and SW lobes, respectively. 

DC07 investigated the influence of an inclination correction on the momentum and concluded that for any inclination, a correction will overestimate the momentum especially in our case where the outflow is oriented almost in the plane of the sky. That is why we will not correct the momentum for inclination but only for missing material close to $\varv_\mathrm{sys}$ by multiplying with 0.46$^{-1}$ (column 10 of Table~3 in DC07), and for hidden atomic material by multiplying with 0.45$^{-1}$ (column 3 of Table~1 in DC07). Considering again an inclination $\alpha$ between $0^\circ$ and 30$^\circ$, the former correction denotes an average value. Then, the momenta increase to $P_\mathrm{corr}=(5.1-23.9)\times 10^{-4}~M_\odot$~km~s$^{-1}$ and $(5.8-14.2)\times 10^{-4}~M_\odot$~km~s$^{-1}$ for the NE and SW lobes, respectively. The values of observed and corrected momentum are listed in Table~\ref{tab:M_CO}.

\subsection{Age and momentum force of the outflow}

Given mass, momentum and projected length of the outflow, we can estimate the dynamical age and determine the momentum force of the outflow. To do so, we use the perpendicular method of DC07 (M5 in vdM13) which, according to the authors, provides the most accurate results compared to other methods introduced in their paper because it has no strong dependence on inclination. Although the perpendicular method is believed to yield the best estimates on the dynamical age and the momentum force of the outflow, there is the more commonly used $\varv_\mathrm{max}$ method \citep[][DC07, M1 in vdM13]{Cabrit92}. We will additionally give the results of this method for a better comparison to the results published in the past for outflows of other young protostars. 

\subsubsection{Perpendicular Method}\label{sec:perp}
The method considers most of the
slow-moving material to be in transverse motion. Hence, we use the
maximum outflow radius $W_\mathrm{lobe}$ measured perpendicular to the jet axis as the characteristic length. We use the channel maps in Fig.~\ref{Appfig} at 5.7~km~s$^{-1}$ and 4.0~km~s$^{-1}$ to measure the maximum radii of the lobes, and find $\sim$330~au and $\sim$200~au for the NE and SW lobes, respectively.
Further, this method uses an intensity-weighted velocity $\left\langle \varv^\prime \right\rangle$ \citep[][DC07]{Cabrit92} which we determine by using Eq.~\ref{eq:mass} and \ref{eq:mom} such that 
\begin{equation}
\langle \varv^\prime\rangle = \frac{P_\mathrm{obs}}{M_\mathrm{obs}}.
\end{equation}
We find $\left\langle \varv^\prime \right\rangle\sim5$~km~s$^{-1}$ for the NE lobe and $\left\langle \varv^\prime \right\rangle\sim2$~km~s$^{-1}$ for the SW lobe.
The values of $W_\mathrm{lobe}$ and $\left\langle \varv^\prime \right\rangle$ are listed in Table~\ref{tab:F_CO}. 
Given these values, we estimate the age of the outflow using the following equation from DC07:
\begin{equation}
t_\mathrm{perp} = \frac{W_\mathrm{lobe}}{3\left\langle \varv^\prime \right\rangle}.
\end{equation}
The results span a range of 100--300~yr depending on the lobe and optical depth. Again, we apply a correction factor to account for projection effects due to the inclination. Table~\ref{tab:F_CO} gives the corrected values for $i=90^\circ$ and 60$^\circ$ using the correction factors giving in column 3 of Table~7 in DC07 (0.83$^{-1}$ for $\alpha=0^\circ$ and 0.64$^{-1}$ for $\alpha=30^\circ$). This correction does not increase the age significantly. The maximum age is now 400~yr.

The observed momentum force is calculated with $F_\mathrm{perp} = P_\mathrm{obs}/t_\mathrm{perp}$ which yields $(0.8-4.6)\times10^{-6}~M_\odot$~km~s$^{-1}$~yr$^{-1}$ and $(0.5-1.6)\times10^{-6}~M_\odot$~km~s$^{-1}$~yr$^{-1}$ for the NE and SW lobes, respectively, depending on the optical depth.
We correct again for hidden atomic material using 0.45$^{-1}$ (column 3 in Table~1 in DC07) and for projection effects, 0.47$^{-1}$ ($\alpha=0^\circ$) and 0.89$^{-1}$ ($\alpha=30^\circ$) (column 5 in Table~7 in DC07). We obtain corrected values of momentum force of $F_\mathrm{corr} = (2.0-21.6)\times10^{-6}~M_\odot$~km~s$^{-1}$~yr$^{-1}$ and $(1.2-7.9)\times10^{-6}~M_\odot$~km~s$^{-1}$~yr$^{-1}$ for the NE and SW lobes, respectively, which we show in Table~\ref{tab:F_CO}.

\subsubsection{$\varv_\mathrm{max}$ method}\label{sec:vmax}

The calculation of the dynamical time requires the maximum, observed velocities  and the length of the lobes. which we give in Table~\ref{tab:props}. Although we determined the velocity for the red- and blueshifted emission in each lobe, we will only use the larger value of the respective lobe as there can only be one maximum velocity which are $\varv^\prime_\mathrm{max}=3.6$~km~s$^{-1}$ and 4.7~km~s$^{-1}$ for the NE and SW lobes, respectively. We determine the dynamical age by dividing the corresponding outflow length $R^\mathrm{proj}_\mathrm{lobe}$ by $\varv^\prime_\mathrm{max}$ which yields 3100~yr and 2500~yr for the NE and SW lobes, respectively. However, these values provide only upper limits since we do not correct for inclination.
Given the age, we compute the momentum force using $F_\mathrm{vmax} = M_\mathrm{obs}\cdot \varv^\prime_\mathrm{max}/t_\mathrm{vmax}$. We do not correct for hidden atomic material to avoid bias since generally, it is not applied for other sources. But following the approach of vdM13, we apply a correction factor from \citet{Cabrit92} which depends on inclination and takes into account material at low velocities. We take the correction factor for $i=70^\circ$ which is the highest inclination given. Hence, we only obtain a lower limit since the outflow seems to be more inclined as shown in Sect.~\ref{sec:props}. \citet{Cabrit92} give uncertainties on their correction factors which we give as an uncertainty on our results. In this way, the momentum force is $F^\mathrm{corr}_\mathrm{vmax} = (0.04-0.13)\times10^{-6}~M_\odot$~km~s$^{-1}$~yr$^{-1}$ and $(0.2-0.4)\times10^{-6}~M_\odot$~km~s$^{-1}$~yr$^{-1}$ for the NE and SW lobes, respectively, depending on optical depth (see Table~\ref{tab:barnard}).

\section{Discussion}\label{sec:discussion}

\subsection{Continuum emission}
The elliptical Gaussian fit performed on the continuum emission map in the
uv plane in Sect.~\ref{ss:continuum} indicates a slight elongation of the 
emission along a direction close to the axis of the outflow, but this 
elongation may not be significant. Given the high inclination of the system, 
seen nearly edge-on, a circumstellar disk would manifest itself with a strong 
elongation perpendicular to the outflow axis. Therefore, we conclude that the 
dust component of diameter $\sim$110~au detected with ALMA must represent the 
inner parts of the collapsing envelope. After rescaling to the new distance of 
192~pc, the mass of the envelope derived by \citet{Belloche11} is 
2.36~$M_\odot$ in a (rescaled) radius of 4800 au. If the envelope has a 
spherical symmetry and the density profile is a power law, then the mass we 
obtain in a radius of $\sim$55~au implies a power-law index of about $-1.5$, 
which is reasonable for a collapsing envelope and strengthens our 
interpretation that the continuum source detected with ALMA corresponds to the 
inner parts of the envelope. If a disk is present, then it must have a radius 
significantly smaller than $\sim$55~au. The unresolved residual component of 2~mJy/beam reported in Sect.~\ref{ss:continuum} may correspond to such a disk, with a mass of $\sim$2$\times 10^{-4}~M_\odot$ if the emission is optically thin but this unresolved 
residual component, as well as the negative ring around, may simply be due to 
our use, for simplicity, of a Gaussian function to fit the envelope emission 
while it may have a power-law structure. The upper limit of 55~au to the disk radius is consistent with the small disks predicted by \citet{Hennebelle16} for early circumstellar disks in collapsing magnetized cores, in which the disk size is self-regulated by magnetic braking and ambipolar diffusion.

\subsection{Episodic ejections}

\begin{figure}
\hspace{-1cm}
\includegraphics[scale=0.33]{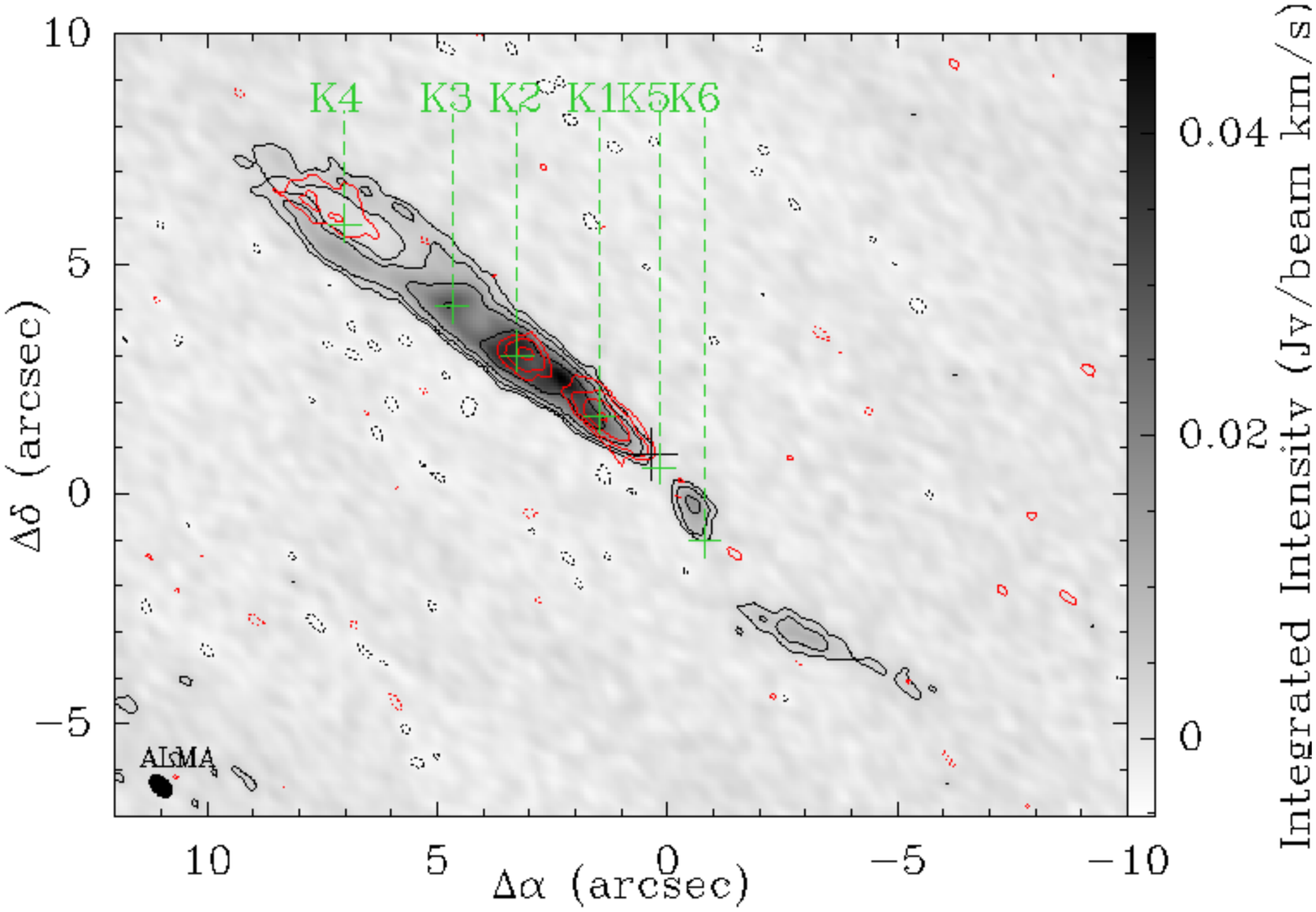}
\caption{Integrated intensity maps toward Cha-MMS1 (black cross) of the high-velocity redshifted CO 3--2 gas [15.0,21.8]~km~s$^{-1}$ shown with red contours and the low-velocity redshifted CO 3--2 gas [9.1,11.6]~km~s$^{-1}$ shown in greyscale and with black contours. The contour levels are $-3\sigma$, 3$\sigma$, 6$\sigma$, 12$\sigma$, and 24$\sigma$ with $\sigma=0.7$~mJy/beam~km~s$^{-1}$ and 1.1~mJy/beam~km~s$^{-1}$ for the emissions at high velocities and low velocities, respectively. The green crosses K1-K6 mark the positions at which the spectra in Figs.~\ref{fig:PV}c-h were taken. The HPBW is shown in the bottom left corner. The black cross marks the position of the continuum peak.}
\label{fig:bullets}
\end{figure}
Figures~\ref{fig:PV} and \ref{fig:bullets} show evidence for the presence of high velocity bullets in the outflow of Cha-MMS1, surrounded by slower-moving material tracing the interface region of the outflow cavity with the ambient medium. The bullets (K1-K6) manifest themselves as compact blobs in Fig.~\ref{fig:bullets}, finger-like elongations in the PV diagrams shown in Figs. 3a-b, and peaks with wings extending to higher velocities in the spectra displayed in Figs.~\ref{fig:PV}c-h. The presence of bullets has been reported in other protostellar outflows and suggests that the ejection phenomenon during the protostellar phase is intermittent in nature \citep[see e.g.,][]{Bachiller96}. The presence of bullets in Cha-MMS1's outflow therefore suggests episodic ejections in this source. \citet{Plunkett15} also inferred intermittency in the outflow of the Class 0 protostar SerpS-MM18 (which they call CARMA~7), and they derived a timescale of ejections of 80--540~yr (not corrected for inclination) based on their PV diagrams. In our case, the knots seem to be separated by $\sim$2$\arcsec$ which translates into a time between two knots of $\sim$20~yr given a distance of 192~pc and assuming a velocity of $\sim$90~km~s$^{-1}$. This is only the case if the knots remain at constant speed. However, in the PV diagrams in Fig.~\ref{fig:PV}, we see that only the bump closest to the centre K1 has this high velocity, K2-K4 move more slowly. \citet{Plunkett15} report a similar behaviour of their outflow which they attribute to an increasing entrainment of and hence, momentum transfer to ambient material leading to the slow-down of the outflow itself.
\citet{Machida14} performed MHD simulations of protostellar outflows and reported on the occurrence of intermittent jets but with more frequent ejections every 1--2~yr. They ascribe the time-variability of the jet to episodic accretion processes, which \citet{Plunkett15} quote as a reason, too.

\subsection{Alignment with the filament}

\citet{Belloche11} show a 870~$\mu$m continuum emission map of the Cha~I region in which Cha-MMS1 is located. Remarkably, the outflow axis is parallel to the filament. Using a sample of 45 sources \citet{Anath08} found that 72\% of protostellar outflows are within 45$^\circ$ of being orthogonal to the filaments in which the sources are embedded. However, based on a slightly larger sample (57 protostars in Perseus), \cite{Stephens17} concluded that the distribution of outflow orientations is more consistent with a random orientation or a mix of parallel and orthogonal angles with respect to the filaments.
Both studies attribute the orientation of the outflows to different core formation processes (see both papers and references therein). 
Since \citet{Stephens17} observed a mix of orientations, they propose a combination of core formation mechanisms to take place in the same cloud.
We conclude from the results of these previous studies that the alignment of the Cha-MMS1 outflow with the filament in which the protostar is embedded is not a seldom phenomenon in star-forming regions.

\begin{figure}
\hspace{-1cm}
\includegraphics[scale=.38]{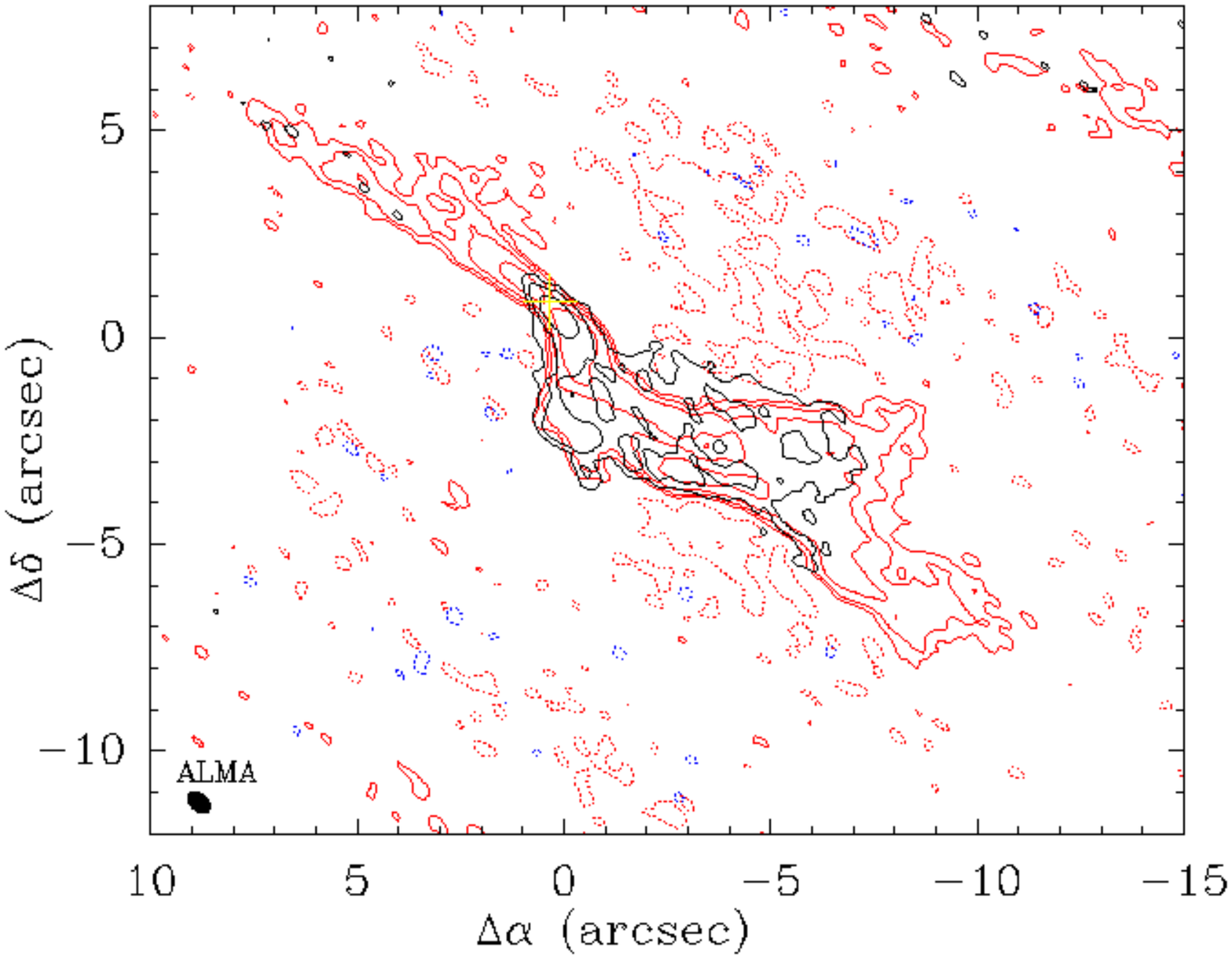}
\caption{Channel map of CO 3--2 emission (red) at 5.7~km~s$^{-1}$ from Fig.~\ref{Appfig} overlaid by $^{13}$CO intensity (black) integrated from 4.9~km~s$^{-1}$ to 5.7~km~s$^{-1}$. The contour levels are -3$\sigma$, 3$\sigma$, 6$\sigma$, 12$\sigma$, 24$\sigma$, 48$\sigma$ with $\sigma=1.8$~mJy/beam and $\sigma=2.5$~mJy/beam for the CO and the $^{13}$CO intensity, respectively. The continuum peak is marked with the yellow cross. The HPBW is shown in the bottom left corner.\label{fig:overlay}}
\end{figure}

\subsection{Spatial extent of the outflow}\label{Sect:extended}

The integrated intensity maps in Fig.~\ref{fig:co_int_map}a have revealed poorly collimated redshifted emission (R1) at the tip of the SW lobe and extended blueshifted emission (B1) which is even separated from the lobe. 
The PV-diagram in Fig.~\ref{fig:PV_ext} reveals that the redshifted emission of the kinked part of the SW lobe has a similar morphology as the redshifted emission in Fig.~\ref{fig:PV}b, while the blueshifted emission seems to be a prolongation of the emission evident at 0--2~$\arcsec$ which presents another indication that B1 belongs to the outflow of Cha-MMS1. Regarding the channel maps in Fig.~\ref{Appfig}, these structures become only prominent at low velocities, that is close to the systemic velocity. The separated blueshifted structure may be part of an older ejection event which has slowed down and broadened with time. Or it may even present a remnant of the generally less collimated outflow occurring during the FHSC phase. Another idea is that the blueshifted emission may
be a bow shock driven by Cha-MMS1 given its orientation and structure.
In any case, if the separated blueshifted emission is part of the Cha-MMS1 outflow, the dynamical age will increase to $t_\mathrm{vmax}\lesssim7400$~yr considering the larger spatial extent of $\sim$3300~au and a maximum velocity at the tip of the outflow of $\varv^\prime_\mathrm{max}\sim2.1$~km~s$^{-1}$ (see Fig.~\ref{Appfig}) which are again not corrected for inclination. On the contrary, $t_\mathrm{perp}$ will not change significantly since the intensity-weighted velocity and the radius of the lobe remain approximately the same.

We cannot rule out that the outflow extends even beyond the field of view observed with ALMA. \citet{Frank14} show how the outflow length of the Class~I protostar B5-IRS1 was found to be much more extended than initially thought once larger maps revealed outflowing emission at farther distances from the source (see their Fig.~9). However, no evidence was found for a NE-SW outflow emanating from Cha-MMS1 in the CO 3--2 single-dish maps of \citet{Belloche06} and \citet{Hiramatsu07}.

\subsection{Lobe misalignment and kink}

In Fig.~\ref{fig:co_int_map}a, we have seen that the axes of the NE and SW lobes have different position angles and that there is a kink within the SW lobe itself.
We compare this result to simulations done by \cite{Ciardi10} where they investigated the influence of a misalignment of the rotation axis to the magnetic field on outflows.
The large-scale magnetic field traced by infrared polarimetry \citep{McGregor94} is roughly orthogonal to the filament in which Cha-MMS1 is embedded. The outflow of Cha-MMS1 is thus roughly perpendicular to this large-scale magnetic field.  
The result presented in Fig.~2c of \citet{Ciardi10} closely resembles the case of the outflow of Cha-MMS1 where the NE lobe is stronger and more collimated than the SW lobe. In this simulation, the rotation axis and the magnetic field are misaligned by 70$^\circ$ which results in an opening angle of the outflow of about 20$^\circ$ similar to what we measure for Cha-MMS1. As a consequence of an increased misalignment, the outflow rate decreases and hence, the protostar mass grows more rapidly. Further, due to precession, also induced by the misalignment, the NE and SW lobes may show different position angles. This is however not clearly visible in Fig.~2c of \cite{Ciardi10}. But they also state, that kinks as well as fragmentation within the outflow lobes can be attributed to this precession. 

These simulations may already explain the morphology of the Cha-MMS1 outflow, however, there may also be another explanation for the observed change in position angle within the SW lobe itself. In a first scenario, the SW lobe may interact with the material traced by the extended continuum emission detected to the Southwest of Cha-MMS1 (see Fig.~\ref{fig:co_int_map}). If this material is dense enough, it could deflect the jet which would result in a change of position angle. The deflection of a jet component was also proposed by \citet{Reipurth2.96} to explain the presence of the HH flow HH~110. Since they could not associate a protostellar source to it, they suggest that this HH object and a second nearby one, HH~270, belong to the same lobe of an outflow driven by the Class~I protostar IRAS~05489+0256 (see their Figs.~3 and 4). When they do this connection, they have to explain a change in position angle and proposed the deflection of the jet by a dense cloud core. \citet{Raga02} performed numerical simulations to test this scenario. They found that a jet with a speed of 300~km~s$^{-1}$ disperses the cloud with which it interacts in less than 1000~yr assuming a cloud-to-jet density of 100. To increase the timescale, they introduce precession of the jet to alternate the impact position and thus, slow-down the destruction of the cloud to a timescale of $\sim$3000~yr. We calculate the density of the continuum where the SW lobe of the Cha-MMS1 outflow changes direction. Given a peak flux density of $4\times 10^{-4}$~Jy$/0.47\arcsec$~beam which translates into an H$_2$ column density of $~\sim$5$\times 10^{22}$~cm$^{-2}$ assuming $\kappa=0.02$~cm$^2$~g$^{-1}$ and an excitation temperature of 10~K, we obtain a mean particle density of 
$\sim$10$^7$~cm$^{-3}$ which should be sufficient to deflect the lobe. Both scenarios described by \citet{Raga02} could apply to Cha-MMS1 since on the one hand, we cannot exclude precession of the outflow, and on the other hand, the Cha-MMS1 outflow may be younger than 1000~yr which means that we would still be able to observe the deflection.

\subsection{The surroundings of Cha-MMS1}\label{sec:neighbours}

Another possibility for the morphology of the SW lobe is that there is a second outflow driven by a different source. Since we find no evidence for another protostar close by which could be responsible for a second outflow, we searched for candidates on larger scales.
We investigated the \textit{Spitzer} data reported by \citet{Luhman08} where several young stellar objects are present (see his Fig.~8). The closest are IRS~2, IRS~4, and the Class~I binary system IRS~6. IRS~6 is approximately located along the direction of the arrow D3 shown in Fig.~\ref{fig:co_int_map}a. If one companion or both drive an outflow whose position angle coincides with that of the kinked SW outflow which we observe, it may show the region of interaction. However, there is no large-scale outflow emission associated with IRS~6, hence, the only possibility may be an atomic outflow turning molecular when interacting with the Cha-MMS1 outflow. However, this is unlikely given the location of IRS~6 in the filament in which Cha-MMS1 is also embedded.

We have shown the outflow emission associated to the Class~0 protostar IRS~4 in Fig.~\ref{fig:co_int_map}c. \citet{Ladd11} discuss the interaction of this outflow with the Cha-MMS1 core. They suggest that the axis of the IRS~4 outflow is almost in the plane of the sky with a wide opening angle which results in blue- and redshifted emission on either side of the central source. This is also evident in the maps of \citet{Hiramatsu07} who show similar maps as \cite{Belloche06} and as in Fig.~\ref{fig:co_int_map}c except that their field of view extends more to the North revealing no clear bipolar structure around IRS~4. In order to link the IRS~4 outflow with the redshifted HH objects HH~49/50 which are located $\sim$0.5~pc to the Southwest of IRS~4 \citep{Ladd11}, they suggest that the lobe passing by Cha-MMS1 interacts with it where only the blueshifted part glances off and gets deflected while the redshifted part continues undisturbed.
Consequently, the blueshifted outflow emission which initially moves southwestward turns to the Southeast (cf. dashed blue contours in Fig.~\ref{fig:co_int_map}c). In that way, the redshifted emission of the IRS~4 outflow may be connected to the redshifted Herbig Haro objects HH~49/50. Before, \citet{Bally06} excluded IRS~4 as the driving source of these HH objects because of the discrepancy of blue- and redshifted emission of the IRS~4 outflow and HH~49/50. In turn, they assigned them to Cha-MMS1. Based on our detection of the outflow and given its position angles, we exclude Cha-MMS1 to be responsible for HH~49/50 but support \citet{Ladd11} who attribute it to IRS~4.

\citet{Ladd11} also debate the influence of the Class~II protostar IRS~2 on the Cha-MMS1 dense core. It may not have a major impact on the Cha-MMS1 outflow but \citet{Ladd11} argue that a wind is pushed by IRS~2 toward Cha-MMS1 compressing the gas and heating it. They also associate the Ced~110 reflection nebula with the Class~II protostar. We think, though IRS~2 seems to illuminate the nebula, its u-shape suggests that it is created by another wind or jet which originates from a source farther in the Northwest. The axis along the assumed direction of motion of this jet or wind through the u-shaped nebula then passes south of IRS~2 and may reach to the SW lobe of the Cha-MMS1 outflow, maybe causing the kink in the SW lobe if it is strong enough.

\cite{Bally06} show large-scale images of the Cha~I region including all HH objects. In their Fig.~4, one sees Cha-MMS1 as well as IRS~4, IRS~6, IRS~2, and several HH objects in their surroundings. Here, it is apparent that HH~49/50 do not coincide with the outflow of Cha-MMS1. In the background, the reflection nebula is evident, as well. According to our suggestion of a more northeasterly located source driving a jet which is responsible for the u-shaped nebula, HH~48 may be a possible candidate. The small-scale structure shown by \citet{Wang06} however, makes HH~48 a less likely candidate due to its orientation.

Furthermore, we want to mention two other objects in this figure of \citet{Bally06} which are HH~929 and HH~936. These two objects are located not far from the prolonged axes of the NE and SW lobes of the Cha-MMS1 outflow, respectively. This means that if the outflow of Cha-MMS1 is indeed spatially more extended than what we see based on our ALMA observations, these two objects might denote the end of a large outflow driven by Cha-MMS1. However, the axes of the NE and the kinked SW lobe miss the locations of the two HH objects by approximately 10$^\circ$, respectively, which means that the outflow must have wandered to a greater extent.

\begin{figure*}
\centering
\includegraphics[width=0.33\textwidth]{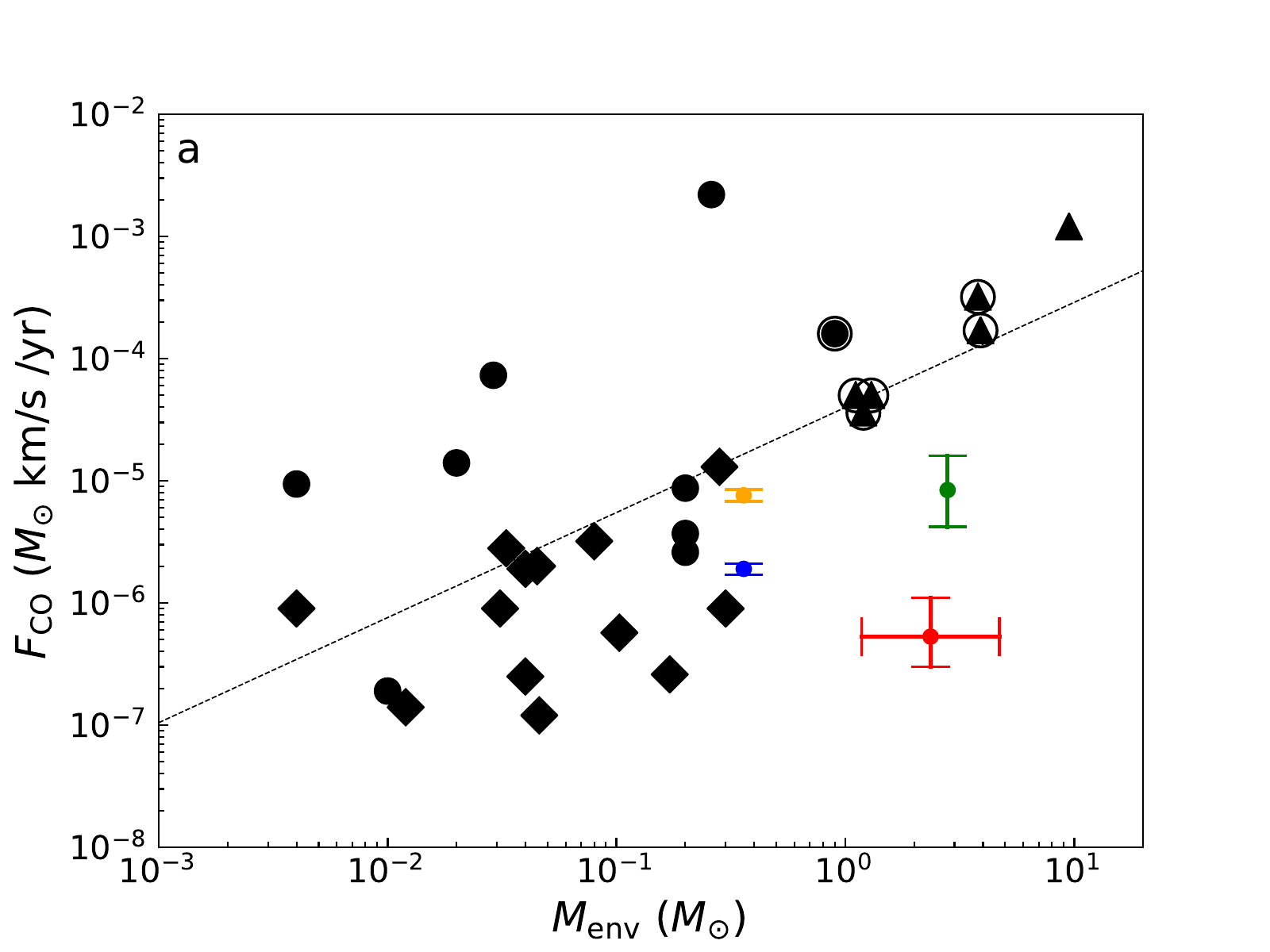}
\includegraphics[width=0.33\textwidth]{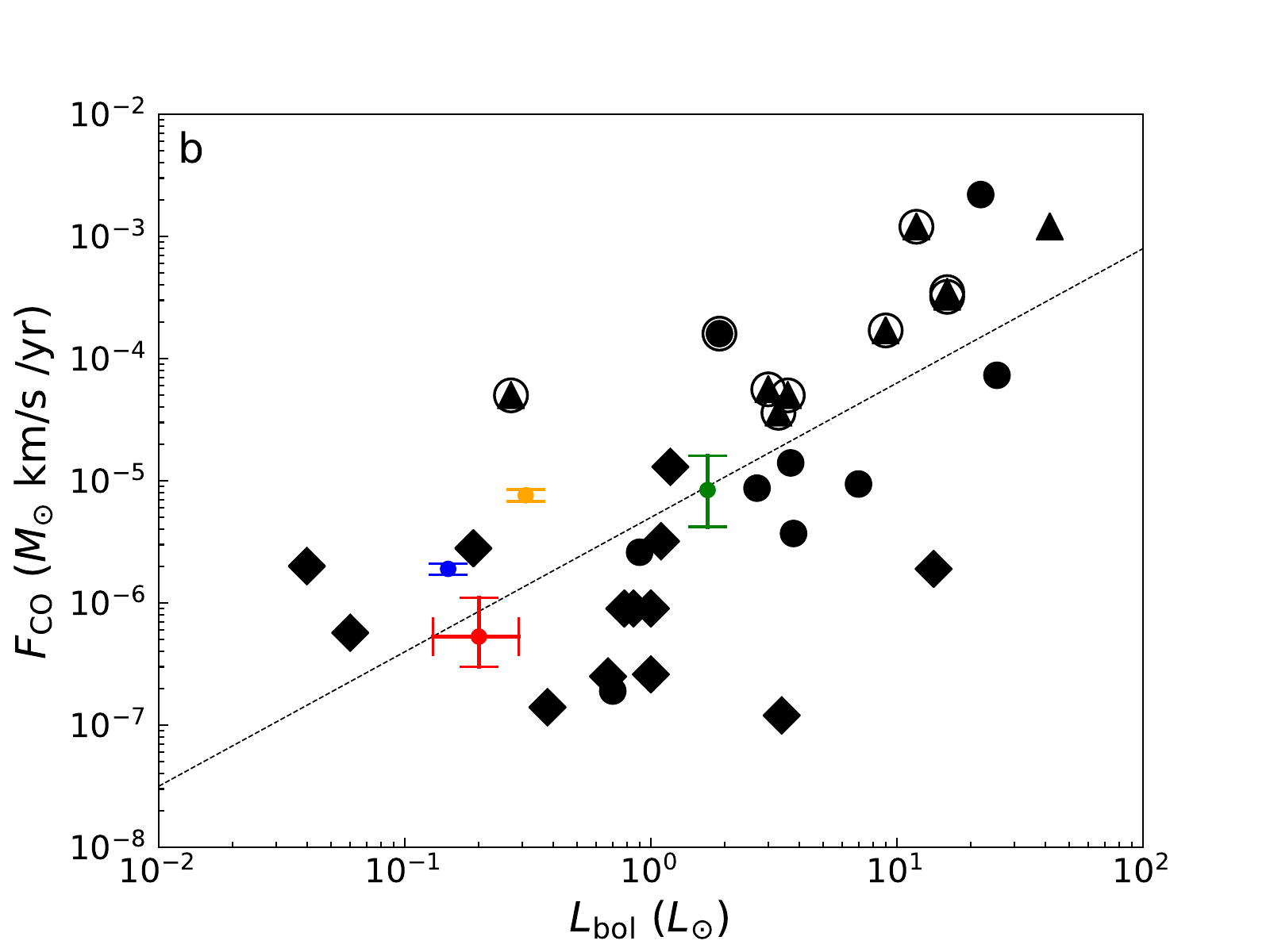}
\includegraphics[width=0.33\textwidth]{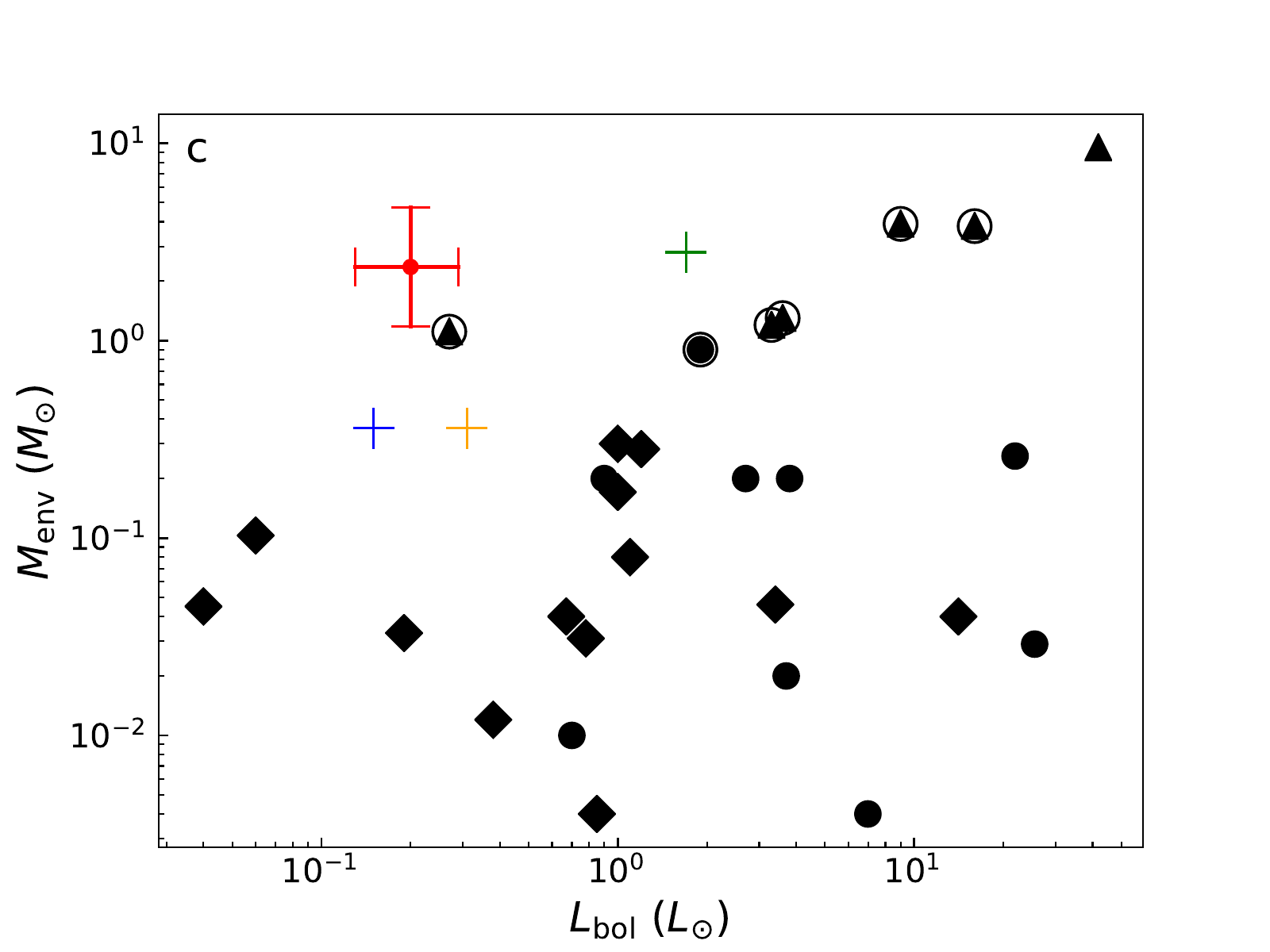}
\caption{Correlation plots for (a) $F_\mathrm{CO}$ and $M_\mathrm{env}$, (b) $F_\mathrm{CO}$ and $L_\mathrm{bol}$, and (c) $M_\mathrm{env}$ and $L_\mathrm{bol}$ adapted from vdM13. Diamond symbols are from their study, triangles from \citet{Cabrit92} and filled circles from \citet{Hogerheijde96}. The Class 0 protostars are encircled. The dashed lines indicate their best fit. The results for Cha-MMS1, GF~9--2 \citep{Furuya19}, B1b-N, and B1b-S \citep{Gerin15} are shown in red, green, blue, and orange, respectively. The error bars on the total momentum force of the outflows represent the uncertainty on the inclination correction adapted from \citet{Cabrit92}. The error bars on the luminosity and the mass of Cha-MMS1 correspond to the range derived by \cite{Tsitali13} and  an uncertainty of a factor of 2 \citep{Belloche11}, respectively. The results for Cha-MMS1 correspond to the optical depth case $\tau_{12}=5$. }
\label{fig:vdM}
\end{figure*}
\begin{table*}
\caption{Outflow properties of B1b-N, B1b-S, GF~9--2, and Cha-MMS1 derived with the $\varv_\mathrm{max}$ method. }
\label{tab:barnard}
\centering 
\begin{tabular}{l l c c c c c c} 
\hline\hline \\[-0.2cm]
Source & Lobe & $\varv^\prime_\mathrm{max}$\tablefootmark{a} & $R^\mathrm{proj}_\mathrm{lobe}$\tablefootmark{b} & $t_\mathrm{vmax}$\tablefootmark{c} & $M_\mathrm{obs}$\tablefootmark{d} & $F^\mathrm{corr}_\mathrm{vmax}$\tablefootmark{e} & $\dot{M}_\mathrm{out}$\tablefootmark{f} \\[0.1cm]
& & (km~s$^{-1}$) & ($10^3$~au) & (kyr) & ($10^{-3}~M_\odot$) & ($10^{-6}~M_\odot$~km~s$^{-1}$~yr$^{-1}$)  & ($10^{-7}M_\odot$~yr$^{-1}$) \\[0.1cm]
\hline\\[-0.3cm]
\multirow{2}{*}{B1b-N\tablefootmark{1}} & Red & 4.0 & 0.6 & 0.7 & 0.4 & 0.7$^{+0.08}_{-0.08}$ & 5.7 \\[0.1cm]
& Blue & 4.5 & 1.3 & 1.4 & 1.2 & 1.2$^{+0.15}_{-0.13}$ & 8.6 \\[0.05cm]
 \hline\\[-0.3cm]
\multirow{2}{*}{B1b-S\tablefootmark{1}} & Red & 6.6 & 3.3 & 2.4 & 1.9 & 2.2$^{+0.27}_{-0.24}$  & 7.9 \\[0.1cm]
 & Blue & 7.6 & 3.0 & 1.9 & 3.0 & 5.4$^{+0.66}_{-0.59}$  & 16 \\[0.05cm]
 \hline\\[-0.3cm]
 \multirow{2}{*}{GF~9--2\tablefootmark{2}} & NE & 3.5 & 2.7 & 3.6 & 0.09 & 0.09$^{+0.09}_{-0.04}$ & 0.3 \\[0.1cm]
 & SW\tablefootmark{*} & 7.8 & 6.0 & 3.6 & 3.4 & 8.3$^{+8.2}_{-4.1}$  & 9.4 \\[0.05cm]
\hline\\[-0.3cm]
\multirow{2}{*}{Cha-MMS1\tablefootmark{**}} & NE & 3.6 & 2.4 & 3.1 & 0.03--0.10 & 0.04$^{+0.04}_{-0.02}$--0.13$^{+0.13}_{-0.06}$  & 0.1--0.3 \\[0.1cm]
& SW & 4.7 & 2.5 & 2.5 & 0.10--0.17 & 0.2$^{+0.2}_{-0.1}$--0.4$^{+0.4}_{-0.2}$  & 0.4--0.7 \\[0.1cm]\hline
\end{tabular}\\[0.3cm]
\tablefoot{\tablefoottext{1}{\citet{Gerin15}.} \tablefoottext{2}{\citet{Furuya19}.}
\tablefoottext{a}{Maximum observed velocity $\varv^\prime_\mathrm{max}=\vert \varv_\mathrm{max}-\varv_\mathrm{sys}\vert$; not corrected for inclination.}\tablefoottext{b}{Projected length of a lobe.} \tablefoottext{c}{Dynamical age $t_\mathrm{vmax}=R^\mathrm{proj}_\mathrm{lobe}/\varv^\prime_\mathrm{max}$.} \tablefoottext{d}{Observed outflow mass.} \tablefoottext{e}{Momentum force $F^\mathrm{corr}_\mathrm{max}=M_\mathrm{obs}\cdot\varv^\prime_\mathrm{max}/t_\mathrm{vmax}$. The error only includes the uncertainty on the correction factor from \citet{Cabrit92}.} \tablefoottext{i}{Outflow rate $\dot{M}_\mathrm{out}=M_\mathrm{obs}/t_\mathrm{vmax}$.}\\ 
\tablefoottext{*}{The SW lobe contains three components for which we report the maximum observed length and velocity and the sum of the masses, momentum forces, and outflow rates.}\\
\tablefoottext{**}{The ranges given for $M_\mathrm{obs}$, $F^\mathrm{corr}\mathrm{vmax}$, and $\dot{M}_\mathrm{out}$ for Cha-MMS1 are based on the two optical depth assumptions for CO ($\tau_{12}<<1$ and $\tau_{12}=5$).}}
\end{table*}

\subsection{Comparison to other Class~0 and Class~I outflows}\label{ss:links}

As there exist various methods of determining the momentum force, results for one source can vary by
more than an order of magnitude (vdM13). Hence, it is rather difficult to make reasonable comparisons
between different sources using results published separately. Former studies \citep[e.g.][vdM13]{Cabrit92, Bontemps96, Hatchell07} however worked out relations between bolometric luminosity $L_\mathrm{bol}$, envelope mass $M_\mathrm{env}$, and momentum force $F_\mathrm{CO}$ which show evolutionary trends from a Class 0 to a Class~I
protostars. It is widely believed that the outflow and the luminosity are both generated by the same
process, that is accretion of matter originating from the surrounding envelope. These correlations make the
three parameters indicators of the evolutionary stage of the protostar. Figure~\ref{fig:vdM} shows
recent results of vdM13 where they combined data from various studies.  Class 0 protostars are encircled. To make a meaningful comparison with the results of vdM13, we have calculated the momentum force using the $\varv_\mathrm{max}$ method and applied the correction factors from \citet{Cabrit92} just as vdM13 in their paper (cf. Sect.~\ref{sec:vmax}).
The results for Cha-MMS1 are shown in red. The envelope mass of Cha-MMS1 $M_\mathrm{env}=2.36~M_\odot$ 
\citep[][corrected for the new distance]{Belloche11} is comparable to those of Class 0
protostars shown in Fig.~\ref{fig:vdM}a. \cite{Belloche11} assume an uncertainty on the mass of at least a factor of 2 which we use as an error bar. The momentum force of the Cha-MMS1 outflow however lies 1--2 orders of magnitude below the momentum forces of these Class 0 protostars, depending on optical depth.
As for the bolometric luminosity, we use the values of \citet{Tsitali13} who worked out the internal luminosity of Cha-MMS1 at a distance of $D=150$~pc and for an inclination of $45^\circ <i< 60^\circ$. We correct their values for the most recent distance of Cha-MMS1 and find $L_\mathrm{int} = 0.13-0.29~L_\odot$. However, it could well be higher considering the higher inclination of Cha-MMS1. From this range, we assume an internal luminosity of $L_\mathrm{int}\approx 0.2~L_\odot$. Cha-MMS1 thus is at the lower end of the Class 0 luminosity range shown in Fig.~\ref{fig:vdM}b where we give the above luminosity range as the uncertainty. Due to its lower luminosity and its envelope mass which is comparable to other Class 0 protostar, Cha-MMS1 manifests itself in the upper left corner in the $M_\mathrm{env}$--$L_\mathrm{bol}$ plot in Fig.~\ref{fig:vdM}c.
The low luminosity may partly be explained by the high inclination when there is a disk of material which is
oriented perpendicular to the jet axis and is efficiently obscuring the central core \citep{Bontemps96}. The rather small values of momentum force may result from Cha-MMS1 being an extremely young Class 0 protostar and maybe, the outflow has not been brought to full power yet
\citep{Smith2000, Hatchell07}. In Sect.~\ref{sec:vmax}, we discussed that the values found for the momentum force using the $\varv_\mathrm{max}$ method only present a lower limit since the inclination seems to be higher than 70$^\circ$ for which we have corrected the results. We have found values which are one to two orders of magnitudes smaller than those found with the perpendicular method in Sect.~\ref{sec:perp}. However, the results for momentum force from the perpendicular method are still lower than momentum forces of other Class~0 protostars.

\subsection{Evolutionary stage of Cha-MMS1}

With ALMA we have been able to detect a bipolar outflow associated with Cha-MMS1 as the driving source
with which we can finally draw a conclusion on the evolutionary stage of the young stellar object. Even without applying any inclination corrections we detect outflow velocities up to $\sim$17~km~s$^{-1}$, well above $\sim$5~km~s$^{-1}$, which is clearly visible in the PV diagram in Fig.~\ref{fig:PV} and the channel maps in Fig.~\ref{Appfig}. This is already inconsistent with FHSC outflow speeds predicted by numerical simulations \citep[e.g.,][]{Commercon12, Machida08} and implies that Cha-MMS1 has already left the
FHSC phase and entered the Class~0 phase. When we correct for inclination we determine maximum outflow velocities as high as at least 90~km~s$^{-1}$ for a likely conservative value of the inclination of 79$^\circ$.

\citet{Commercon12} also showed that a first core can be detected at 24~$\mu$m and 70~$\mu$m if
the source is inclined by less than 60$^\circ$ to the line of sight. However, if a source at the FHSC
stage is inclined by more than 60$^\circ$ it cannot be detected at these wavelengths over its whole
lifetime since most of the radiation emanating from the first core is reprocessed by the surrounding
envelope. We find that the axis Cha-MMS1's outflow lies almost in the plane of the sky implying that it should not be
observable at 24~$\mu$m and 70~$\mu$m if it were in the FHSC stage.  The fact that \cite{Belloche06} detected Cha-MMS1 at these wavelengths supports the presumption of it not being a FHSC but in
a more evolved state, that is a Class~0 protostar.  

In Sect.~\ref{sect:intro} we have introduced the dense cores B1b-N and B1b-S where the former seems to
be a FHSC while the latter is proposed to be already in a more evolved state \citep{Gerin15}. The
properties of the outflows are summarised in Table~\ref{tab:barnard} where, based on the observed properties found by \citet{Gerin15}, we compute the momentum force in the same way as we have done it for Cha-MMS1 in Sect.~\ref{sec:vmax} except that we use the correction factor from \citet{Cabrit92} for $i=50^\circ$ since the blue- and redshifted outflow emission of B1b-N and B1b-S are well separated. The
momentum forces of B1b-N and B1b-N are larger than those found for Cha-MMS1. Their ranges of results are indicated in blue (B1b-N)
and orange (B1b-S) in Fig.~\ref{fig:vdM}. The total observed outflow mass of Cha-MMS1 is smaller by
an order of magnitude compared to both, B1b-N and B1b-S. Based on the $\varv_\mathrm{max}$ method, the dynamical age of the Cha-MMS1 outflow seems to be comparable to that of B1b-S while being slightly higher compared to B1b-N. 
The internal luminosity $L_\mathrm{int}=0.2~L_\odot$ of Cha-MMS1 lies in between the luminosities of B1b-N and B1b-S (0.15~$L_\odot$ and 0.31~$L_\odot$) while the envelope mass is higher compared to that of B1b (0.36~$M_\odot$) \citep{Hirano14}. \\
\\Another recent study by \citet{Furuya19} reports on the detection of an outflow driven by the Class~0
protostar GF~9--2. \citet{Furuya19} determine the outflow properties using a distance of 200~pc but more recent results suggest a larger distance of 474~pc \citep[][C.~Zucker priv.~comm.]{Zucker19}. We rescale their results and show them in Table~\ref{tab:barnard}. After rescaling, the SW lobe of the GF~9-2 outflow shows the largest spatial extent in this sample. The dynamical age is only slightly higher compared to Cha-MMS1.
\citet{Furuya19} uses two excitation temperatures to derive the outflow mass where we will compare to the values at $T=22.6$~K. Though the mass of the NE lobe is significantly smaller than that of the SW lobe, the total mass of the GF~9-2 outflow is roughly an order of magnitude higher than the mass of the Cha-MMS1 outflow. 
We determine the momentum force using the $\varv_\mathrm{max}$ method and apply the correction factor from \citet{Cabrit92} for $i=70^\circ$ to the results since \citet{Furuya19} find an inclination of $\sim$65$^\circ$ from SED modelling. The results for GF~9--2 are shown in green in Fig.~\ref{fig:vdM}. The momentum force of the GF~9-2 outflow is roughly an order of magnitude higher than the value obtained for the outflow of Cha-MMS1. 
However, they assume only optically thin emission and due to the ambiguous morphology of the outflow, it could also be inclined
as highly as Cha-MMS1 which would in principle increase the momentum force. In Fig.~\ref{fig:vdM}, we use the internal luminosity and envelope mass adopted by \citet{Maury19} from \citet{Wiesemeyer97} for a distance of 200~pc, $\sim$0.3~$L_\odot$ and $\sim$0.5~$M_\odot$, and rescaled them to the new distance (1.7~$L_\odot$ and 2.8~$M_\odot$). 

All three
sources show rather small values of momentum force compared to the other Class~0 protostars in
Fig.~\ref{fig:vdM}. This may be the result of the sources being extremely young Class~0 protostars (or a FHSC in the case of B1b-N) so their outflows have not come to full power, yet.  \citet{Bontemps96} suggests an outflow rate of $\sim$10$^{-6}~M_\odot$~yr$^{-1}$ for young Class~0 protostars. All sources introduced here show a lower rate of a few times 10$^{-8}~M_\odot$~yr$^{-1}$ to a few times 10$^{-7}~M_\odot$~yr$^{-1}$ which may support the hypothesis that their outflows are not yet fully developed. 

Admittedly, the derived properties of the Cha-MMS1 outflow contradict the predictions of a FHSC but agree with those of a young Class~0 protostar. The comparison to other young Class~0 protostars and a promising FHSC candidate supports the classification of Cha-MMS1 as a young Class~0 protostar rather than a FHSC.

\section{Conclusion}\label{sec:conclusion}

We observed Cha-MMS1 at high angular resolution in CO 3--2 and $^{13}$CO 3--2 emission and in continuum emission at $\sim$900~$\mu$m. The goal of the project was to search for a slow small-scale
outflow to probe if Cha-MMS1 is at the stage of the first hydrostatic core. The ALMA observations reveal a bipolar outflow clearly associated with Cha-MMS1 as the driving
source. Our main results are the following:

\begin{enumerate}
 \item The continuum emission detected at small scale originates from the inner parts of the Cha-MMS1 protostellar envelope. No direct evidence for a disk was found and we derived an upper limit to the disk radius of 55~au.
 \item The two lobes are oriented in the northeast southwest direction and both show 
 blue- as well as redshifted emission indicating that the system is seen nearly edge-on. Given its high inclination, Cha-MMS1 should not be visible at 24~$\mu$m if it were a FHSC according to MHD simulations, while it was detected by \textit{Spitzer} at this wavelength.
 \item In the spectra, we observe maximum velocities of $\sim$16.4~km~s$^{-1}$ and $\sim$10.6~km~s$^{-1}$ in the NE and SW lobes, respectively, which even without correction for inclination are already inconsistent with the low velocities ($<$5~km~s$^{-1}$) predicted by MHD simulations for FHSC outflows. 
 \item We find an observed outflow mass of 3$\times 10^{-4}~M_\odot$ which is at least an order of magnitude lower compared to other Class~0 protostars. However, correcting for hidden atomic material and material at low velocities, the mass increases by a factor $\sim$4.5. 
 \item We determine the dynamical age and momentum force of the outflow using two different methods. With the perpendicular method, we obtain ages of $\sim$200~yr while with the $\varv_\mathrm{max}$ method we get an upper limit of $\sim$3000~yr. We derive a total momentum force of
$(1.6-3) \times 10^{-5}~M_\odot$~km~s$^{-1}$~yr$^{-1}$ using the perpendicular method which is approximately two orders of magnitude higher than the value obtained with the $\varv_\mathrm{max}$ method which is $\sim 5\times 10^{-7}~M_\odot~$km~s$^{-1}$~yr$^{-1}$. 
 \item Compared to outflows of more evolved Class~0 protostars, the outflow of Cha-MMS1 has a much smaller momentum force. However, the result is similar to momentum forces of protostars with similar dynamical age, such as GF~9--2 or B1b-S (and B1b-N), suggesting that these young Class~0 protostars have not yet brought their outflows to full power. 
\end{enumerate}

Based on these characteristics, we conclude that Cha-MMS1 is not a FHSC but has already entered the Class~0 protostar phase. Assuming that the outflow is not more extended than what we observe with ALMA, the outflow may be one of the youngest ever observed with a dynamical age of 200--3000~yr.

\begin{acknowledgements}
We thank Anastasia Tsitali for her contribution during the initial stage of this project.    
\end{acknowledgements}

\begin{appendix}

\section{PV diagram of R1 and B1}
\begin{figure}[h!]
\vspace{-0.5cm}
    \centering
    \includegraphics[width=0.46\textwidth]{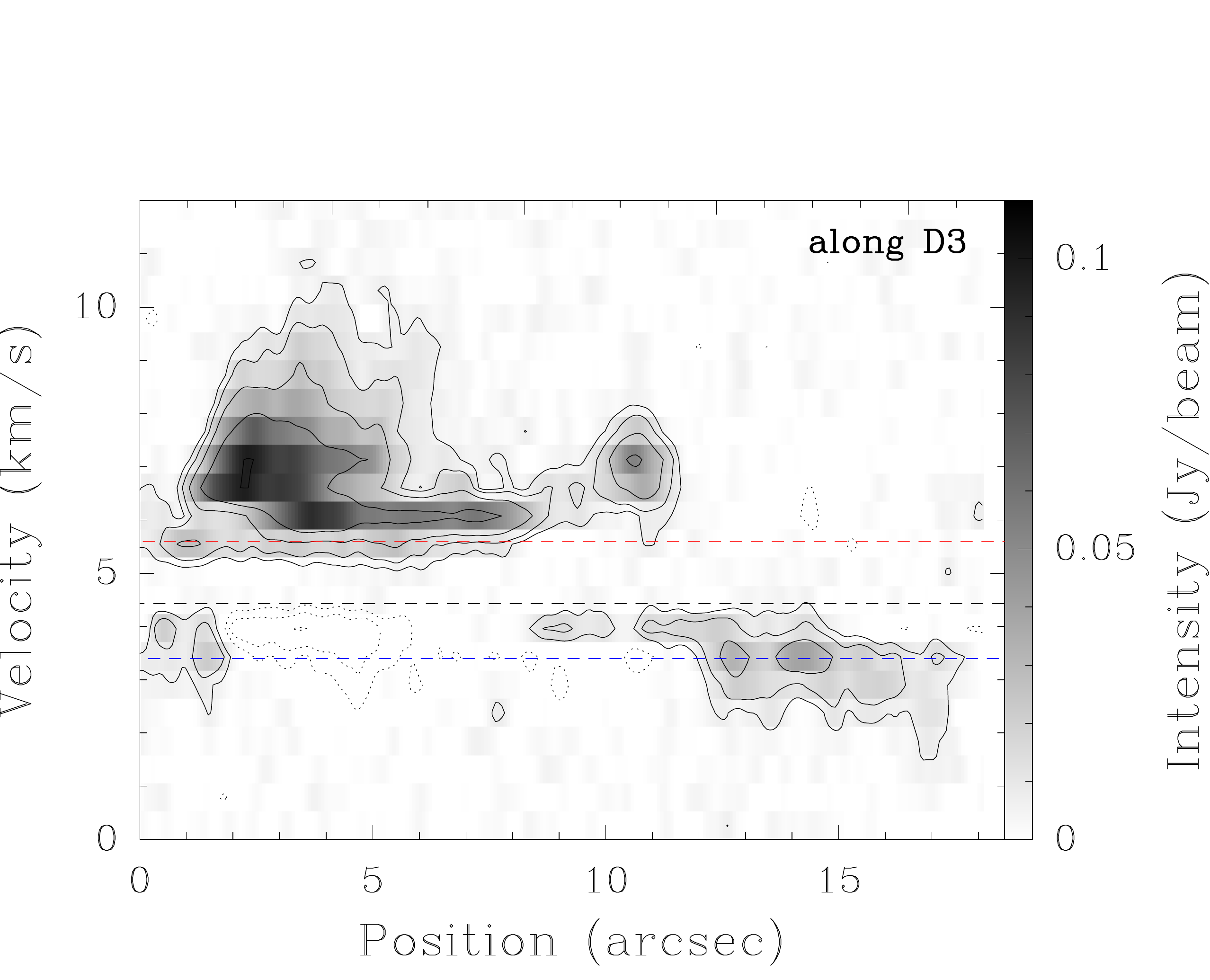}
    \caption{CO 3--2 position-velocity diagram along the arrow labeled D3 shown as a green dashed arrow in Fig.~\ref{fig:co_int_map}a. The position axis originates from the start of the arrow in Fig.~\ref{fig:co_int_map}a. The intensity is smoothed over 5 channels. The contours are $-12\sigma$, $-6\sigma$, $-3\sigma$, $3\sigma$, $6\sigma$, $12\sigma$, $24\sigma$, and $48\sigma$ with $\sigma=2.0$~mJy/beam. The red and blue dashed lines show the inner integration limits while the black dashed line indicates the systemic velocity of Cha-MMS1. }
    \label{fig:PV_ext}
\end{figure}
\section{Uncertainties}
\subsection{Inclination}
The overlap between the red- and blueshifted emission suggests that the outflow of Cha-MMS1 has an axis close to the plane of the sky with an inclination $i$ between 79$^\circ$ and 101$^\circ$. 
However, it is also possible that the inclination is smaller than our results suggest. The channel maps in Fig.~\ref{Appfig} and \ref{Appfig2} indicate that the outflow becomes broader at velocities close to the systemic velocity. In Fig.~\ref{fig:overlay} this indication is evident, too. Here, we overlay the CO emission of the channel at 5.7~km~s$^{-1}$ taken from the channel maps in Fig.~\ref{Appfig} and the $^{13}$CO emission integrated in the range [4.9,5.7]~km~s$^{-1}$. This broadening would result in a larger opening angle and hence, increase the range of possible inclinations. This increase in opening angle with decreasing velocity has also been seen in early studies of other protostellar outflows \citep[e.g.,][]{Bachiller90}. However, in the case of Cha-MMS1, this broadening is only seen after the kink noticed in the SW lobe, so it may result from the process which generates this kink and does not trace the intrinsic property oft the outflow as driven by Cha-MMS1. This is the reason why we did not take this broad width as the characteristic width of the outflow to estimate its inclination in Sect.~\ref{sec:props}.

\subsection{Momentum force}

We calculated the dynamical time and the momentum force of the outflow using two different methods where the results obtained from the perpendicular method are higher than those obtained from the $\varv_\mathrm{max}$ method by approximately an order of magnitude.
Based on the discussion in DC07, we claim that the perpendicular method gives the more reliable estimates because the required parameters, which are the intensity-weighted velocity and the radius of the lobe, depend only minorly on inclination. On the contrary, the maximum outflow velocity and the length along the axis of the outflow as used in the $\varv_\mathrm{max}$ method strongly depend on inclination. To account for the inclination in the $\varv_\mathrm{max}$ method, we applied a correction from \cite{Cabrit92} for $i=70^\circ$ because this is the highest inclination for which they give the correction factor. Hence, we obtain only a lower limit for the momentum force. Further, in order to determine the maximum velocity for the $\varv_\mathrm{max}$ method, one needs to evaluate whether the highest velocities really trace the bulk of the outflow or bullets.

\subsection{Opacity and outflow mass}

Both methods consider the outflow mass which we derived from integrated intensities. The conversion factor $K$ depends on the excitation temperature for which we assume 30~K. However, it would change by less than a factor of 2 for temperatures of 50~K or 20~K. Whatever method is used, another uncertainty results from the optical depth. To account for the high optical depths at the inner velocity cut-offs, we use the $^{13}$CO emission until a certain threshold and correct for optical depth wherever it is possible\footnote{We tested that using only $^{12}$CO yields masses and momenta underestimated by factors 1.1--5 depending on the lobe.}. We do not determine a correction for channels in which the intensity ratio of the isotopologues is smaller than 1, which may underestimate the intensity in these channels. Beyond the $^{13}$CO intensity threshold, we add the $^{12}$CO emission where we consider two cases, optically thick ($\tau_\mathrm{12}=5$) and thin ($\tau_\mathrm{12}<<1$) CO emission. In that way, the results span a range in which
the true mass and momentum force of the outflow should lie. The maximum recoverable scale of our observations is $\sim$7$\arcsec$. The 
outflow lobes are significantly narrower than this ($\sim$2--3$\arcsec$) so
the outflow emission is not expected to be much affected by spatial filtering.
$^{12}$CO outflow emission close to the systemic velocity of the source is
likely hidden behind optically thick, larger-scale envelope emission filtered 
out by the interferometer, and thus cannot trace the outflow well, but our 
strategy to use $^{13}$CO in the cases where $^{12}$CO is optically thick 
should reduce the impact of this on our estimates of the outflow mass and 
momentum force. 
Additionally, the mass and momentum force may still be underestimated when molecular dissociation is not considered (DC07). Only in the case of the perpendicular method, we allow for more undetected atomic gas by applying the correction factor introduced by DC07. We do not apply this correction to the results from the $\varv_\mathrm{max}$ method, in order to make more reliable comparison to other sources for which these factor was not applied either. 

Further, independently of the methods, we integrate intensities within fixed velocity intervals. Though we use the $^{13}$CO emission to account for material closer to $\varv_\mathrm{sys}$ and though we apply a correction factor to take into account missing material at velocities close to $\varv_\mathrm{sys}$, we may still miss some material at low velocities. Figure~\ref{fig:overlay} shows that the outflow emission gets less collimated at low velocities. 
However, the choice of the outer velocity cut-offs does not introduce large uncertainties to the integrated intensities since most of the emission is found at lower velocities. 

\subsection{Outflow radius}

In Fig.~\ref{fig:overlay} we have seen that the outflow is less collimated at low velocities which results in a larger outflow radius $W_\mathrm{lobe}$. We determined $W_\mathrm{lobe}$ at the lowest possible velocity showing the largest radii. But if the true radius is even larger, the quantities computed from the perpendicular method of DC07, that is dynamical age $t_\mathrm{perp}$ and momentum force $F_\mathrm{perp}$, will increase and decrease, respectively.  

\subsection{Contamination from IRS~4}

In the special case of Cha-MMS1, there is emission originating from the nearby Class~I protostar
Ced~110~IRS~4 and its outflow which might contaminate the emission of CO 3--2 
\citep{Belloche06, Hiramatsu07, Tsitali13}. We have shown the outflow emission of IRS~4 in Fig.~\ref{fig:co_int_map}c and the  larger-scale emission at velocities close to $\varv_\mathrm{sys}$ probably tracing ambient material of the filament in Fig.~\ref{fig:co_int_map}d. While the emission of the IRS~4 outflow may in principle contaminate our observations, most of the large-scale emission appears to be filtered by ALMA and the morphology of the emission detected close to Cha-MMS1 is undoubtedly originating from Cha-MMS1 itself.

\onecolumn
\section{CO and $^{13}$CO channel maps}

\begin{figure*}[h!]
 \centering
 \includegraphics[width=1.\textwidth]{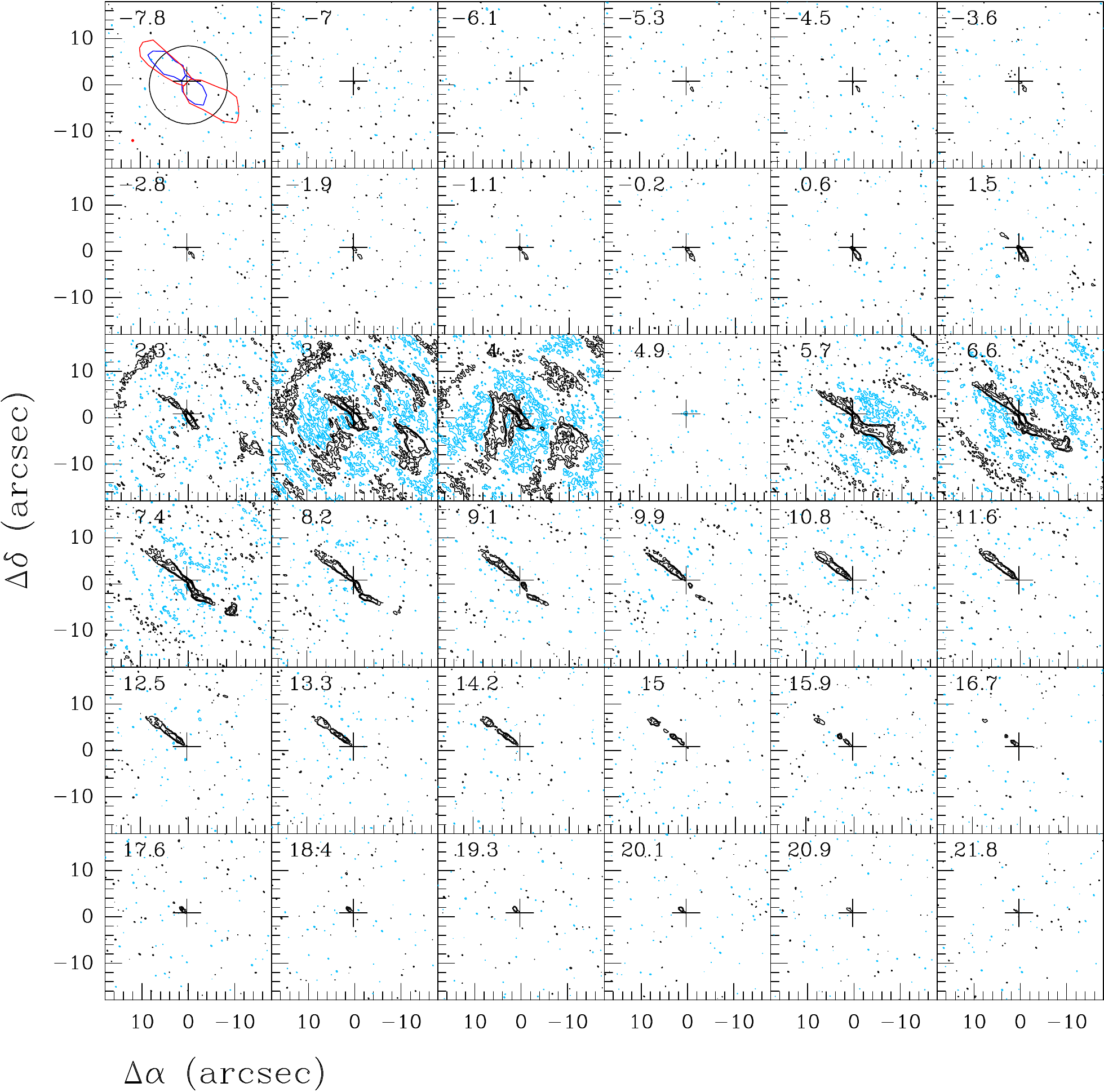}
 \caption{Velocity-channel maps of the CO 3--2 emission toward Cha-MMS1. The cross marks the peak continuum position. Each map is averaged over eight channels corresponding to $\sim0.88$~km~s$^{-1}$ with the central velocity shown in km~s$^{-1}$ in the upper left corner. The contour steps are $-12\sigma$, $-6\sigma$, $-3\sigma$, $3\sigma$, $6\sigma$, $12\sigma$, $24\sigma$, and $48\sigma$ with $\sigma = 1.8$~mJy/beam. The light blue contours represent the negative levels. The primary beam (black circle) and the synthesized beam (red ellipse) are shown in the upper left panel. The red and blue contours represent the masks used for the mass calculation  (see Sect.~\ref{ss:mass}). The maps are not corrected for primary beam attenuation. The systemic velocity is $\varv_\mathrm{sys}=4.43$~km~s$^{-1}$ \citep{Belloche06}.}
 \label{Appfig}
\end{figure*}

\begin{figure*}[h!]
 \includegraphics[scale=1.]{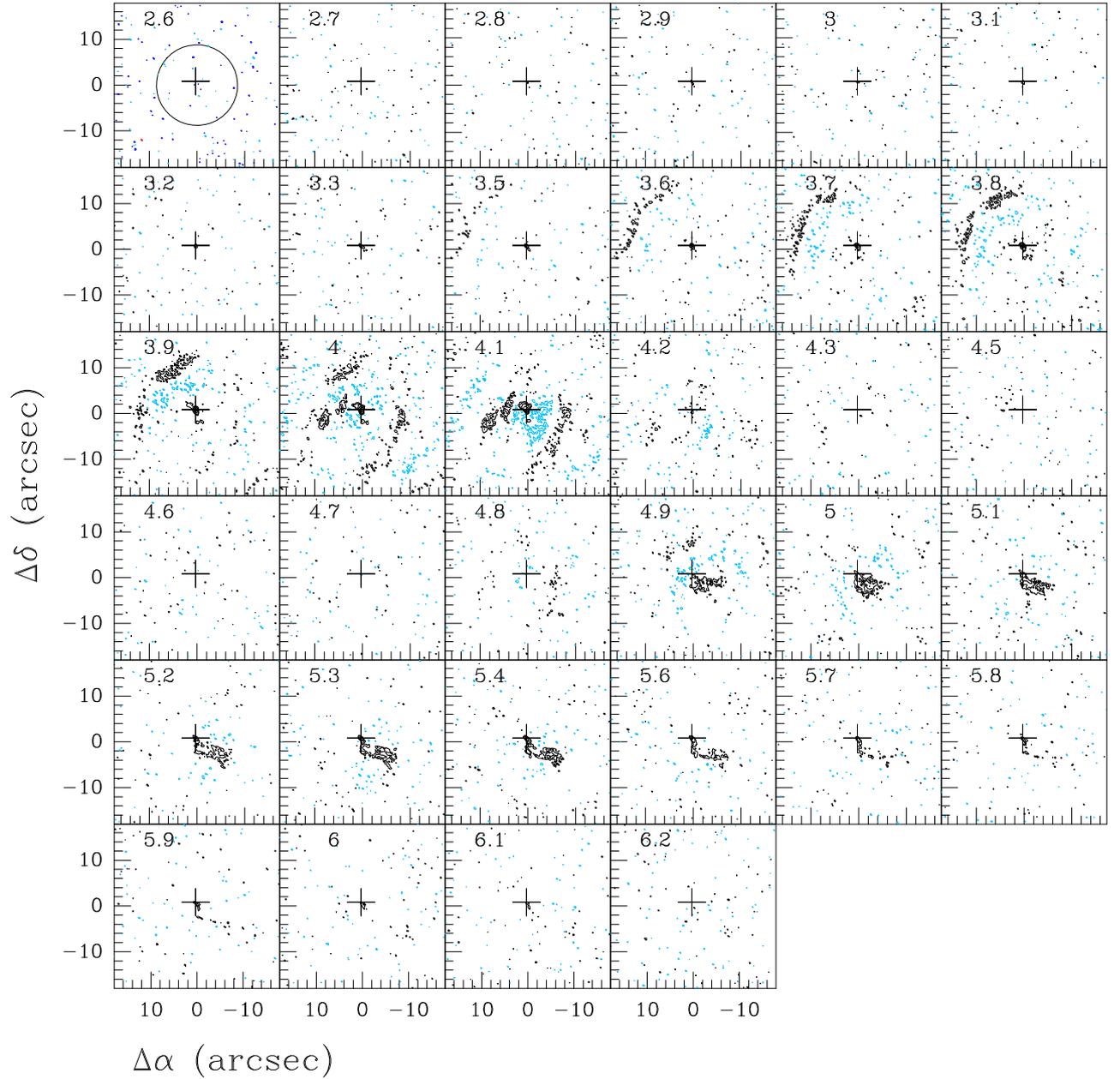}
 \caption{Velocity-channel maps of the $^{13}$CO 3--2 emission toward Cha-MMS1. The cross marks the peak continuum position. The central velocity is shown in km~s$^{-1}$ in the upper left corner. The contour steps are $-6\sigma$, $-3\sigma$, $3\sigma$, $6\sigma$, $12\sigma$, $24\sigma$, and $48\sigma$ with $\sigma = 4.7$~mJy/beam. The light blue contours represent the negative levels. The primary beam (black circle) and the synthesized beam (red ellipse) are shown in the upper left panel. The maps are not corrected for primary beam attenuation. The systemic velocity is $\varv_\mathrm{sys}=4.43$~km~s$^{-1}$ \citep{Belloche06}.}
\label{Appfig2}
\end{figure*}

Figure~\ref{Appfig} shows the CO 3--2 channel maps smoothed over 8 channels (0.88~km~s$^{-1}$) which cover a velocity range from 22.6~km~s$^{-1}$ to $-7$~km~s$^{-1}$. 
Figure~\ref{Appfig2} shows the $^{13}$CO 3--2 channel maps which cover a velocity range from 2.6~km~s$^{-1}$ to 6.2~km~s$^{-1}$. 

\newpage
\end{appendix}
 
\end{document}